\documentclass[journal,twoside]{IEEEtran}
\usepackage{cite}
\usepackage{amsmath,amssymb,amsfonts}
\usepackage{algorithmic}
\usepackage{graphicx}
\usepackage{textcomp}
\usepackage{booktabs}
\usepackage{url}

\begin{document}
\title{Super-Resolved Microbubble Localization in Single-Channel Ultrasound RF Signals Using Deep Learning}
\author{Nathan Blanken, Jelmer M. Wolterink, Hervé Delingette, Christoph Brune, Michel Versluis, and Guillaume Lajoinie
\thanks{\copyright 2022 IEEE.  Personal use of this material is permitted.  Permission from IEEE must be obtained for all other uses, in any current or future media, including reprinting/republishing this material for advertising or promotional purposes, creating new collective works, for resale or redistribution to servers or lists, or reuse of any copyrighted component of this work in other works.}
\thanks{N.B., J.M.W., C.B., M.V., and G.L. acknowledge funding from the 4TU Precision Medicine program supported by High Tech for a Sustainable Future, a framework commissioned by the four Universities of Technology of the Netherlands.}
\thanks{N. Blanken, M. Versluis, and G. Lajoinie are with the Physics of Fluids group, MESA+ Institute for Nanotechnology, Technical Medical (TechMed) Centre, University of Twente, Enschede, The Netherlands (e-mail: n.blanken@utwente.nl; m.versluis@utwente.nl; g.p.r.lajoinie@utwente.nl).}
\thanks{J. M. Wolterink and C. Brune are with Applied Mathematics, Technical Medical (TechMed) Centre, University of Twente, Enschede, The Netherlands (e-mail: j.m.wolterink@utwente.nl; c.brune@utwente.nl).}
\thanks{Hervé Delingette is with the EPIONE team, INRIA, Sophia Antipolis, France. (e-mail: herve.delingette@inria.fr)}
\thanks{The code used for this work is available at \protect\url{https://github.com/MIAGroupUT/SRML-1D}.}}

\maketitle

\begin{abstract}
Recently, super-resolution ultrasound imaging with ultrasound localization microscopy (\textsc{ulm}) has received much attention. However, \textsc{ulm} relies on low concentrations of microbubbles in the blood vessels, ultimately resulting in long acquisition times. Here, we present an alternative super-resolution approach, based on direct deconvolution of single-channel ultrasound radio-frequency (\textsc{rf}) signals with a one-dimensional dilated convolutional neural network (\textsc{cnn}). This work focuses on low-frequency ultrasound (1.7 MHz) for deep imaging (10 cm) of a dense cloud of monodisperse microbubbles (up to 1000 microbubbles in the measurement volume, corresponding to an average echo overlap of 94\%). Data are generated with a simulator that uses a large range of acoustic pressures (5-250 kPa) and captures the full, nonlinear response of resonant, lipid-coated microbubbles. The network is trained with a novel dual-loss function, which features elements of both a classification loss and a regression loss and improves the detection-localization characteristics of the output. Whereas imposing a localization tolerance of 0 yields poor detection metrics, imposing a localization tolerance corresponding to 4\% of the wavelength yields a precision and recall of both 0.90. Furthermore, the detection improves with increasing acoustic pressure and deteriorates with increasing microbubble density. The potential of the presented approach to super-resolution ultrasound imaging is demonstrated with a delay-and-sum reconstruction with deconvolved element data. The resulting image shows an order-of-magnitude gain in axial resolution compared to a delay-and-sum reconstruction with unprocessed element data.
\end{abstract}


\vspace{30pt}


\section{Introduction}
\label{sec:introduction}

\IEEEPARstart{U}{ltrasound} is one of the most commonly used medical imaging modalities. It is non-invasive, real-time, safe, and, compared to \textsc{ct} and \textsc{mri}, portable and inexpensive~\cite{Szabo2004}. In addition, ultrasound can quantify physical quantities such as tissue elasticity~\cite{Sigrist2017} and blood flow~\cite{Ricci2018}. However, blood is a poor ultrasound scatterer, resulting in low-contrast images. To improve contrast, microbubbles, which behave like (nonlinear) ultrasound scatterers~\cite{Overvelde2010}, can be injected into the bloodstream~\cite{Engelhard2018}. Current clinical practice uses polydisperse microbubbles, i.e. microbubbles with a broad size distribution (typically 1-10~~µm). Due to the direct relation between resonance frequency and bubble radius, most bubbles will be driven off-resonance, resulting in a weak response. Recent advances in the fabrication and characterization of monodisperse microbubbles (i.e. all of the same size) allow for exploiting their nonlinear behavior in an unprecedented way~\cite{Segers2018b,Helbert2020}, providing even better tissue contrast and signal-to-noise ratio. Despite these advances in contrast imaging, the resolution of ultrasound images remains diffraction-limited, i.e. limited to the wavelength of the transmitted pulse, which typically ranges from 100 µm to 1 mm. The diffraction limit hinders the detection of small lesions, especially in deep tissue, into which short-wavelength ultrasound cannot penetrate. 

Inspired by successes in optical microscopy~\cite{Huang2009}, researchers have devised methods to overcome the diffraction limit in ultrasound imaging~\cite{Errico2015,Christensen-Jeffries2020,Couture2018,Bar-Zion2018,Piepenbrock2019,Sloun2019}. The effort has mostly been focused on ultrasound localization microscopy (\textsc{ulm})~\cite{Errico2015,Christensen-Jeffries2020,Couture2018}. In \textsc{ulm}, the microvasculature is imaged by tracking the microbubbles that are confined within the capillaries. \textsc{ulm} requires a sparse distribution of microbubbles in the microvasculature, which ensures the separation of their point spread functions.  By tracking the centroids of these point spread functions as the microbubbles flow through the capillaries, the microvasculature and its flow field (magnitude and direction) are mapped.

Although \textsc{ulm} has achieved unprecedented in-vivo image quality, its application in the clinic is challenging for several reasons. Firstly, \textsc{ulm} requires low microbubble concentrations in the microvasculature. Therefore, many ultrasound images are required to map the microvessels, making real-time imaging difficult to achieve: mapping an entire capillary network can take tens of minutes~\cite{Hingot2019}. Secondly, \textsc{ulm} often relies on beamformed envelope-detected data (reconstructed images). Consequently, the quality of the super-resolved image depends on the beamforming algorithm. For high-resolution images, beamforming may suffer from speed-of-sound inhomogeneity. Thirdly, beamformed images lack the phase and frequency information that is present in the raw radio-frequency (\textsc{rf}) transducer element data. In particular, the phase of the microbubble response depends on its size~\cite{Sijl2011}, resulting in an unknown relation between the centroid of the grayscale point spread function and the bubble location, especially for polydisperse microbubbles. 

Multiple efforts have been made to go beyond these limitations~\cite{Bar-Zion2018,Piepenbrock2019}, including an image-based deep-learning-based approach~\cite{Sloun2019}. Recently, a deep-learning method was reported that does not rely on beamformed data and performs detection and localization of linear scatterers directly from element data~\cite{Youn2020}. This method allows for detecting concentrated scatterers with a localization uncertainty an order of magnitude smaller than the wavelength. Since the article considers linear scatterers, the approach is only valid for microbubbles driven at high frequencies. For deep imaging, low-frequency ultrasound is required, at which microbubbles behave nonlinearly.

Here, we develop a deep-learning approach to achieve super-resolution from raw transducer element \textsc{rf} data acquired from a randomly distributed monodisperse microbubble cloud (radius 2.4~µm). We simulate the full, nonlinear microbubble response, allowing us to focus on deep imaging (up to 10~cm) with low-frequency ultrasound (1.7~MHz), which is near the microbubble resonance frequency (the resonance frequency depends on the acoustic pressure~\cite{Overvelde2010,Helfield2019}, which ranges from 5 to 250~kPa). Simulated data gives direct access to the ground truth and allows for the rapid generation of a large, diverse data set. With the simulated data, we train a one-dimensional (1D) dilated convolutional neural network (\textsc{cnn}) to deconvolve single-channel \textsc{rf} signals from raw element data, i.e. \textit{before} beamforming. We demonstrate detection and localization of microbubbles in \textsc{rf} signals with dense microbubble distributions.


\section{Methods}\label{SectionMethods}

A single-channel \textsc{rf} signal $V(t)$ obtained from a microbubble cloud can be represented as a convolution of the temporal microbubble distribution $\phi(t) = \sum_n\delta(t-t_n)$ with the echo of a single microbubble $V_\mathrm{MB}(t)$: $V(t) = \phi(t)*V_{\mathrm{MB}}(t)$, where $t_n$ is the arrival time of the $n$-th microbubble echo. However, solving the inverse problem, i.e. recovering $\phi(t)$ from $V(t)$, is complicated by the variability of the convolution kernel $V_\mathrm{MB}$. Not only the \textit{amplitude} of the convolution kernel varies, e.g. due to acoustic attenuation or simply due to a differently set transmit voltage, but also its \textit{shape}, due to nonlinear wave propagation and the nonlinear microbubble response. Variations in the properties of individual bubbles, most importantly their radii, also increase the variability of the convolution kernel. Kernel variability impedes the use of analytical deconvolution strategies. Instead, we use a 1D convolution neural network (\textsc{cnn}) to deconvolve the 1D \textsc{rf} signals. 

\begin{figure}[ht]
    \centering
    \includegraphics[width=\columnwidth]{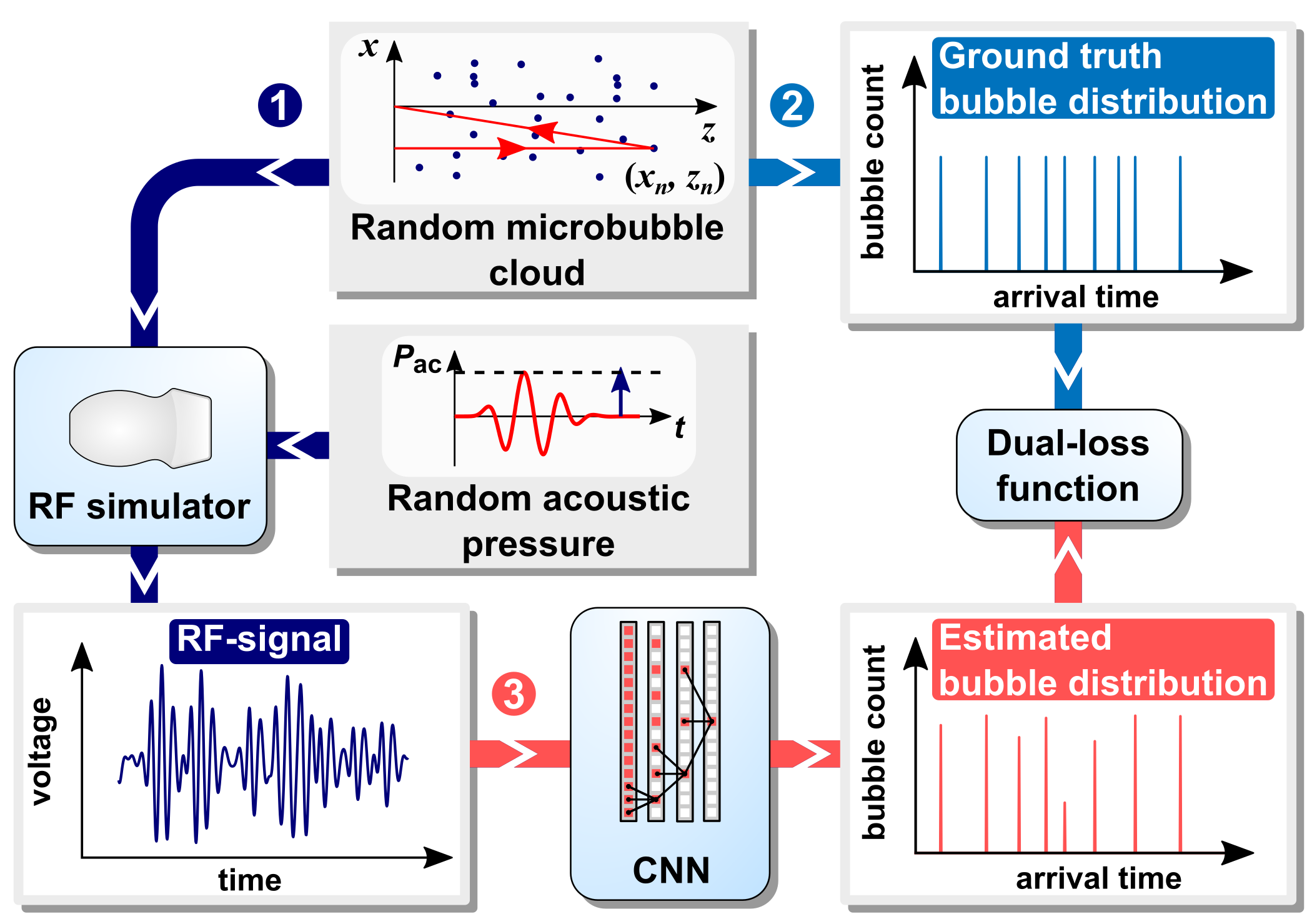}
    \caption{\textbf{Super-resolution strategy.} \textbf{1} A randomly distributed microbubble cloud (random number of bubbles and random coordinates) serves as input to a simulator that generates one-dimensional (1D) \textsc{rf} signals. The individual bubble radii are drawn from a narrow (monodisperse) distribution. The acoustic pressure amplitude is also randomly selected. \textbf{2} The bubble coordinates are also used to compute the 1D ground truth distributions (arrival times of bubble echoes). \textbf{3} A 1D dilated convolutional neural network (\textsc{cnn}) is trained with a dual-loss function to detect and localize microbubbles in an \textsc{rf} signal.}
    \label{blanke1}
\end{figure}

Mathematically, the network needs to map $\Lambda: U\rightarrow \varphi$. Here, $\varphi$ and $U$ are discrete-time (i.e. sampled) representations of the temporal microbubble distribution $\phi(t) = \sum_n\delta(t-t_n)$ and of the \textsc{rf} signal $V(t)$, respectively. The network is trained on \textsc{rf} signals generated with a simulator developed in-house. The training strategy is summarized in Fig.~\ref{blanke1}. The number of bubbles $N_\mathrm{MB}$, the lateral and axial coordinates ($x_n$ and $z_n$) of each bubble, the initial radius $R_0$ of each bubble (drawn from a narrow size distribution), and the acoustic pressure amplitude $P_\mathrm{ac,max}$ of the transmit pulse are randomized for each \textsc{rf} signal simulation. The radius and the acoustic pressure are randomized to produce a realistic kernel variability. We only include \textsc{rf} signals received by the central transducer element, so that each \textsc{rf} signal corresponds to a different randomly assigned bubble distribution and acoustic pressure.

To simplify the simulations, we consider the transmitted wave as an ideal plane wave. Although the wave is invariant in the lateral ($x$) direction, the amplitude and frequency content of the wave are not constant along the propagation ($z$) axis, due to attenuation and nonlinear propagation. Furthermore, we put the bubbles in water, a medium with a homogeneous and known speed of sound~$c$. Therefore, the relation between the bubble coordinates $(x_n,z_n)$ and the echo arrival times $t_n$ is simply (see Fig.~\ref{blanke1}, step 2):
\begin{equation}\label{EqArrivalTimes}
t_n = (z_n + \sqrt{x_n^2 + z_n^2})/c
\end{equation}
This relation is used to construct the ground truth distributions $\varphi$. The ground truth $\varphi$ has the same length as the input signal $U$, allowing for a point-to-point prediction. To compare the predictions of the network to the ground truth, we use a novel dual-loss function that exhibits both properties of a classification loss and a regression loss. 

Super-resolution is often expressed as a fraction of the wavelength $\lambda$. Since we work in the time domain, we introduce a related \textit{diffraction time}, $t_\lambda$. We define $t_\lambda = 2\lambda/c$, which is the arrival time difference of two echoes coming from scatterers separated by one wavelength in the axial direction. The diffraction time is approximately the duration of the ultrasound pulse and is therefore closely linked to the temporal point spread function (the microbubble echo). To allow for sufficient super-resolution, we set the sampling interval of $U$ and $\varphi$ to $\Delta t=16$~ns. $\Delta t/t_\lambda = 0.014$, which is far below current super-resolution limits.

\subsection{RF signal simulator}\label{SectionMethodsSimulator}

\subsubsection{Geometry and transmit pulse}\label{SectionGeometry}
\begin{figure}[ht]
    \centering
    \includegraphics[width=\columnwidth]{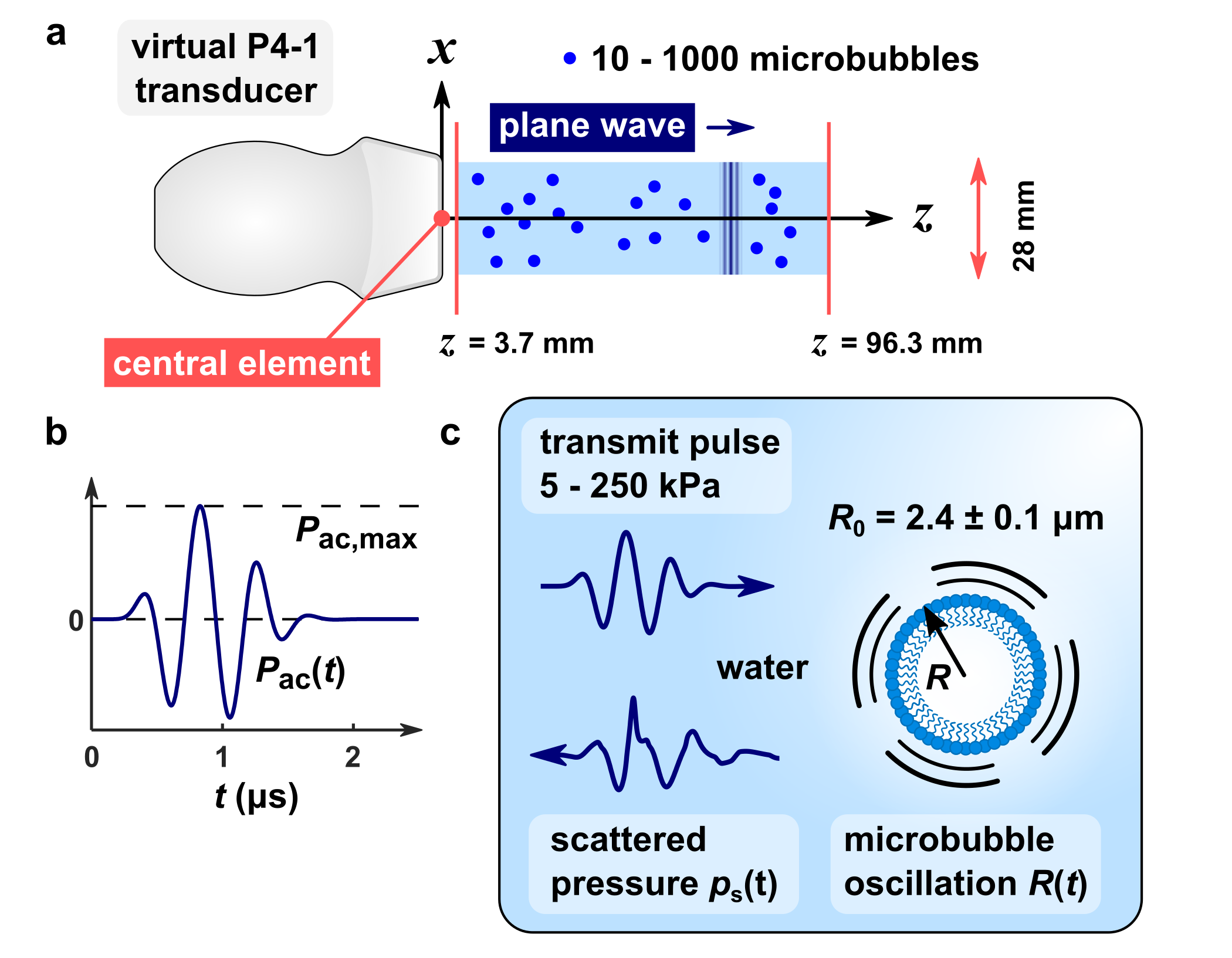}
    \caption{\textbf{RF signal simulator.} \textbf{a} For each simulation, the number of microbubbles is randomly selected: $N_\mathrm{MB}\sim\mathcal{U}\{10,1000\}$. The bubbles are randomly distributed in a two-dimensional domain. A virtual P4-1 transducer emits a short-pulse plane wave which excites the bubbles.  All bubble echoes arriving at the center element are added up and convolved with the receive impulse response of the transducer to produce an \textsc{rf} signal. \textbf{b} Pulse shape. For each simulation, the acoustic pressure is randomly selected: $P_\mathrm{ac,max}\sim\mathcal{U}(5~\mathrm{kPa},~250~\mathrm{kPa})$. \textbf{c} Each bubble is assigned a random radius: $R_0\sim\mathcal{N}\left(2.4~\textrm{µm}, (0.14~\textrm{µm})^2 \right)$. The radial oscillation $R(t)$ is computed by solving a Rayleigh-Plesset equation. From  $R(t)$, the scattered pressure is computed.}
    \label{blanke2}
\end{figure}

For each simulation, we randomly distribute monodisperse microbubbles in a two-dimensional (2D) domain (Fig.~\ref{blanke2}a) that is 28~mm wide and extends from $z=3.7$~mm to $z=96.3$~mm in depth. The imaging depth is set to 100~mm, but no bubbles are placed beyond $z = 96.3$~mm to ensure that the echoes of bubbles near the end of the domain are fully captured. Furthermore, no bubbles are placed in front of $z=3.7$~mm to prevent the signal from growing without bound near the transducer surface due to the 1/r decay of the scattered signals, where $r$ is the distance between the bubble and the receiving element. Since the speed of sound $c$ is 1480~m/s, bubble echoes arrive at the central transducer element between 5~and 131~µs after the start of the transmit pulse (Eq.~\ref{EqArrivalTimes}). 

The number of microbubbles $N_\mathrm{MB}$ in each simulation is randomly drawn from a uniform distribution between 10 and 1000: $N_\mathrm{MB}\sim\mathcal{U}\{10,1000\}$. These numbers represent an increasing degree of point spread function (microbubble echo) overlap. Since a bubble echo is about 2~µs long,  $N_\mathrm{MB}=10$ corresponds to an average echo spacing of 13~µs, and the echoes, statistically, rarely overlap. Conversely, with $N_\mathrm{MB}=1000$, the average spacing is 0.13~µs, corresponding to an average overlap of 94\% between two adjacent echoes. Nevertheless, for $N_\mathrm{MB}=1000$, the average minimum distance between bubbles is 0.8~mm, which is more than 300 times  the bubble radius, allowing bubble-bubble interactions to be neglected~\cite{Segers2018}.  

We model the transmit and the receive characteristics of the transducer upon those of a P4-1 transducer (Philips ATL), see supplementary material. Given the limited transducer bandwidth (-6~dB receive bandwidth 1.4-3.6~MHz) we set the driving frequency to 1.7~MHz to ensure that both the fundamental and the second harmonic bubble response can be recorded. A 1-cycle pulse drives the transducer and is convolved with the transmit impulse response of the transducer. Fig.~\ref{blanke2}b shows the shape of the resulting transmit pulse $P_\mathrm{ac}(t)$. The transmit pulse, normalized by its maximum, is rescaled for each simulation by a randomly selected acoustic pressure $P_\mathrm{ac,max}\sim\mathcal{U}(5~\mathrm{kPa},~250~\mathrm{kPa})$. 

\subsubsection{Nonlinear wave propagation}
Ultrasound propagates nonlinearly in water. Since we argue that higher harmonics are an important feature in our data, we incorporate nonlinear propagation into our simulator.  For an ideal plane wave (1D), we can use the Burgers equation. In water, the attenuation (in Np/m) scales as $\alpha(\omega)=\alpha_0\omega^2$, where $\omega=2\pi f$ is the angular frequency. Therefore, the Burgers equation can be written as~\cite{Blackstock1985,Szabo2004}:
\begin{equation}
    \frac{\partial p}{\partial z} = \frac{\beta}{\rho_\mathrm{L}c^3}p\frac{\partial p}{\partial \tau} + \alpha_0 \frac{\partial^2 p}{\partial \tau^2},
\end{equation}
with $\beta$ the nonlinearity parameter, $\rho_\mathrm{L}$ the density of the liquid and $\tau$ the retarded time $\tau = t - z/c$.
Let $p_0(\tau)$ be the pressure at the transducer surface. We write $p(\tau) = p_1 + p_2 + ...$, with $p_1 >> p_2$. To first order, $p_1(z,\tau) = \mathcal{F}^{-1}\{\mathcal{F}\{p_0(\tau)\}\mathrm{exp}(-\alpha_0\omega^2z)\}$, where $\mathcal{F}$ and $\mathcal{F}^{-1}$ denote the temporal Fourier transform and its inverse, respectively. To second order, the solution can be approximated by (see supplementary material):
\begin{equation}
    p_2(z,\tau) = \mathcal{F}^{-1}\left\{ B(\omega)\frac{e^{-2\alpha_0\omega_0^2z}  - e^{-\alpha_0\omega^2z}}{\alpha_0\omega^2 - 2\alpha_0\omega_0^2}\right\}.
\end{equation}
with $B(\omega) = \beta/(\rho_\mathrm{L}c^3)\mathcal{F}\{p_0\partial p_0/\partial\tau\}$. The third and higher harmonics are not computed, as these fall outside the bandwidth of the transducer. The resulting local acoustic pressure wave $P_\mathrm{ac} = p_1 + p_2$ drives the microbubbles.

\subsubsection{Microbubble response}
We compute the evolution of the bubble radius $R(t)$ using \textsc{matlab}'s ode45 solver to solve the Rayleigh-Plesset equation as reported in~\cite{Marmottant2005}:
\begin{multline}\label{Marmottant}
    \rho_\mathrm{L}\left(R\ddot{R} + \frac{3}{2}\dot{R}^2\right) = \left[P_0 + \frac{2\sigma(R_0)}{R_0}\right]\left(\frac{R}{R_0}\right)^{-3\kappa}\left(1 - \frac{3\kappa}{c}\dot{R}\right) \\- P_0 - \frac{2\sigma(R)}{R} - \frac{4\mu\dot{R}}{R} - \frac{4\kappa_\mathrm{s}\dot{R}}{R^2} - P_\mathrm{ac}(t).
\end{multline}
Here, $R_0$ is the initial bubble radius, $P_0$ is the ambient pressure, $\kappa$ is the polytropic exponent of the gas, $c$ is the speed of sound in the liquid, $\kappa_\mathrm{s}$ is the shell viscosity, and $P_\mathrm{ac}(t)$ is the acoustic pressure wave at the location of the microbubble. The function $\sigma(R)$ is the nonlinear surface tension of a coated microbubble, which contributes substantially to the nonlinear character of the bubble. For $\sigma(R)$, we use an experimentally measured surface tension curve~\cite{Segers2018}. The quantity $\mu$ is a parameter that accounts for both viscous and thermal damping: $\mu = \mu_\mathrm{vis} + \mu_\mathrm{th}$, where $\mu_\mathrm{vis}$ is the dynamic viscosity of the liquid and $\mu_\mathrm{th}$ is a thermal damping parameter. $\mu_\mathrm{th}$ and the polytropic exponent $\kappa$ are calculated with a linear theory of bubble oscillations by Prosperetti~\cite{Prosperetti1977}. $\kappa_\mathrm{s}$ is a shell damping parameter~\cite{Segers2018}. An overview of the simulation parameters is given in the supplementary material.

With Eq.~\ref{Marmottant}, we found that bubbles with $R_0=2.4$~µm resonate at 1.7~MHz when driven at 75~kPa. A typical polydispersity index (standard deviation divided by mean radius) is 5\%~\cite{Segers2018}. Therefore, the initial radius of each bubble is randomly drawn from the Gaussian distribution $R_0\sim\mathcal{N}\left(2.4~\textrm{µm}, (0.14~\textrm{µm})^2 \right)$.

\subsubsection{RF signal construction}
Knowing $R(t)$ from solving Eq.~\ref{Marmottant}, the scattered pressure can be calculated with~\cite{Keller1956}:
\begin{equation}
    p_\mathrm{s}(r,\tau) = \frac{\rho_\mathrm{L}R(\tau)}{r}\left(R(\tau)\ddot{R}(\tau) + 2\dot{R}^2(\tau)\right),
\end{equation}
Note that we have dropped the rapidly decaying $r^{-3}$ near-field term from the original equation and that $p_\mathrm{s}$ and $R$ are again a function of the retarded time $\tau = t - r/c$, although now in spherical coordinates. To account for the attenuation of the scattered wave, we modify the scattered pressure (see supplementary material) as follows:
$p_\mathrm{s,att}(r,\tau) = \mathcal{F}^{-1}\{\mathcal{F}\{p_\mathrm{s}(r,\tau)\}\mathrm{exp}(-\alpha_0\omega^2r)\}$.
We must note that the acoustic attenuation in water is low~\cite{Azhari2010}, and the last equation is a minor correction.

The pressure signal sensed by the receiving transducer element is constructed by adding up the scattered pressures from all microbubbles. To compute the final \textsc{rf} signal (voltage-time curve), the pressure signal is convolved with the impulse response of the virtual transducer.

\subsubsection{Data set generation}
To construct the ground truth distributions, we represent the temporal bubble distribution $\phi(t) = \sum_n\delta(t-t_n)$ as an integer-valued array $\mathbf{\varphi} = (\varphi_1,\varphi_2,...,\varphi_N)$ with the same sampling interval (16~ns) as the \textsc{rf} signal.  The arrival times $t_n$ are calculated with Eq.~\ref{EqArrivalTimes}. For each bubble, $t_n$ is rounded off to the nearest time sample $i\Delta t$ and $\varphi_i$ is incremented by 1. The resulting array $\varphi$ gives the bubble count per grid point in the \textsc{rf} signal. We used our simulator to generate 5000 $U-\varphi$ pairs that were split into separate data sets for training, validation, and testing (model evaluation).

\begin{figure}[ht]
    \centering
    \includegraphics[width=\columnwidth]{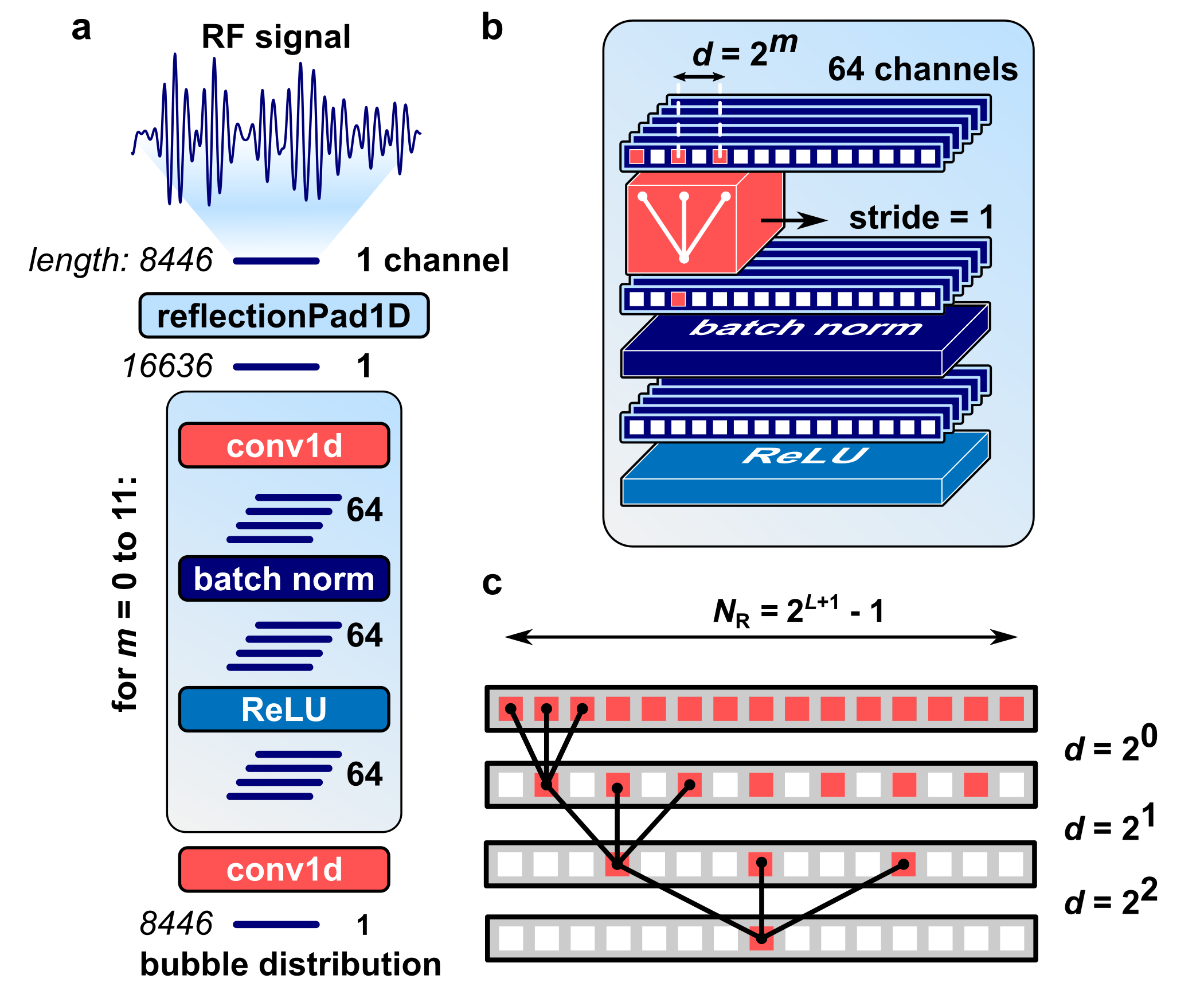}
    \caption{\textbf{Architecture of the dilated convolutional neural network.} \textbf{a} Architecture overview, showing a 1D reflection padding layer, 1D convolutional layers, batch normalization layers and Rectified Linear Unit (ReLU) activation layers. \textbf{b} Detailed illustration of the three-layer block outlined in \textbf{a} for $m=1$, showing the convolution kernel, the stride, and the dilation rate $d$. \textbf{c}. Stack of convolutional layers illustrating the exponential expansion of the receptive field $N_R$ as a function of the number of convolutional layers $L$.}
    \label{blanke3}
\end{figure}

\subsection{Network architecture}\label{SectionMethodsArchitecture}

The receptive field of the convolutional neural network, i.e. the section of the input that can affect one point in the output, needs to be large. For example, the response of a single microbubble lasts approximately 2~µs, corresponding to 125 grid points in an \textsc{rf} signal. To capture a single bubble response, the receptive field should be at least of that order. To aggregate a larger-scale context, an even larger receptive field is required. Whereas large receptive fields in e.g. image analysis are often achieved through encoder-decoder architectures~\cite{Ronneberger2015}, such architectures do not preserve translational equivariance and cannot naturally be applied to the signal lengths that we use. Instead, we use dilated convolution kernels to exponentially expand the receptive field of the network with a linear increase in the number of parameters~\cite{Yu2016,Oord2016}.

In a dilated \textsc{cnn}, the dilation rate $d$, i.e. the spacing between the kernel elements (Fig.~\ref{blanke3}b), increases with increasing network depth: $d = 2^m$, with $m$ the layer index, starting from $m=0$. Provided a stride of 1 is used in all convolutions, such a dilation rate prevents loss of resolution or coverage~\cite{Yu2016}. Fig.~\ref{blanke3}c shows that the receptive field of a dilated \textsc{cnn} with kernel size $k=3$ and $L$ convolutional layers is $N_\mathrm{r} = 2^{L+1}-1$. With only $L=12$, $N_\mathrm{r} = 8191$, which is almost the entire length of the input signal. By contrast, the receptive field of an undilated \textsc{cnn} ($d=1$ for all layers) with $k=3$ is $N_\mathrm{r} = 2L + 1$. To achieve $N_\mathrm{r} = 8191$, an undilated \textsc{cnn} would need 4095 convolutional layers. Training such a deep network is impractical. 

Fig.~\ref{blanke3}a shows the full network architecture. The reflection padding layer increases the signal length from 8446 to 16636. Each convolutional layer reduces the signal length by $2d$. The reflection padding layer assures that the output has the same length as the ground truth distributions (which equals the length of the input signals). The first convolutional layer connects a single input channel to 64 channels. All intermediate convolutional layers  connect 64 channels to 64 channels. Each convolutional layer is followed by a batch normalization layer (to speed up training)~\cite{Ioffe2015} and a Rectified Linear Unit (ReLU) activation layer. The final layer is a linear output layer (a convolutional layer with kernel size 1), which merges the 64 channels into one output channel that corresponds to the microbubble distribution.

\begin{figure}[ht]
    \centering
    \includegraphics[width=\columnwidth]{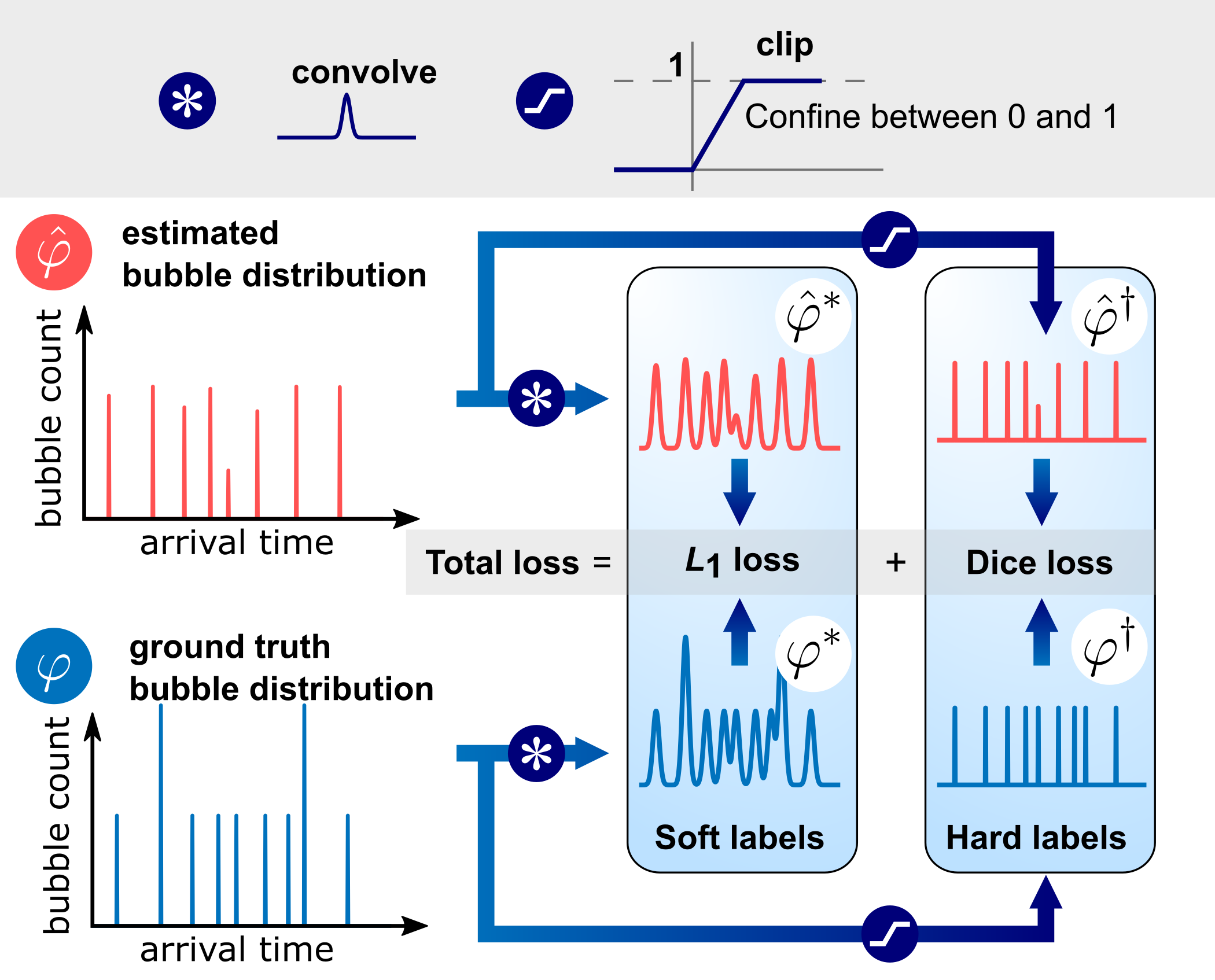}
    \caption{\textbf{Dual-loss function}. The ground truth label $\varphi$ is convolved with a Gaussian kernel to generate a soft label $\varphi^*$. In parallel, $\varphi$ is clipped to produce a binary label $\varphi^\dagger$. The same operations are applied to the network output $\hat{\varphi}$ to generate a soft prediction $\hat{\varphi}^*$ and a binary prediction $\hat{\varphi}^\dagger$. The total loss is a linear combination of an $L_1$ loss, applied to $\varphi^*$ and $\hat{\varphi}^*$, and a Dice loss, applied to $\varphi^\dagger$ and $\hat{\varphi}^\dagger$.}
    \label{blanke4}
\end{figure}

\subsection{Model training}\label{SectionDualLoss}

\subsubsection{Microbubble localization as classification: hard label training}
The bubble count array $\varphi$ is a sparse distribution. It has a length of $N = 8446$ grid points, of which 7864 can be nonzero (because the bubbles are only located between $z = 3.7$~mm and $z=96.3$~mm). For each simulation, 10-1000 microbubbles are distributed over these 7864 grid points. Consequently, the probability of two or more bubbles sharing the same grid point is low. In fact, for super-resolved imaging of microbubbles, the bubble count at a particular grid point is irrelevant, only the presence of a bubble matters. Therefore, the sparse, integer-valued array $\varphi$ can be represented as a binary array $\varphi^\dagger$, with elements valued 1 (positive, one or more bubbles) or 0 (negative, no bubble). This reduces the retrieval of the microbubble distribution to a binary classification problem. However, the sparsity of $\varphi^\dagger$ results in class imbalance: the number of negatives greatly exceeds the number of positives. A binary loss function that can handle class imbalance is the (soft) Dice loss~\cite{Milletari2016,Sudre2017}:
\begin{equation}
    \mathrm{DL} = 1 - \frac{2\sum_{i=1}^N \hat{\varphi}^\dagger_i \varphi^\dagger_i}{\sum_{i=1}^N \hat{\varphi}^\dagger_i + \sum_{i=1}^N \varphi^\dagger_i},
\end{equation}
where $\varphi^\dagger$ represents the ground truth, $\hat{\varphi}^\dagger$ the predicted binary microbubble array and $N$ the total number of grid points. Note that $\hat{\varphi}^\dagger_i$ can take any value between 0 and 1 and DL is minimized if $\hat{\varphi}^\dagger = \varphi^\dagger$.

\subsubsection{Microbubble localization as regression: soft label training}\label{SectionSoftLoss}
A drawback of the Dice loss is that it compares the ground truth $\varphi^\dagger$ to the prediction $\hat{\varphi}^\dagger$ on a per-grid-point basis. Consequently, there is no \textit{localization tolerance}: a predicted bubble that is only one grid point away from a ground truth bubble counts as a false positive and increases the loss value. Unless the network can achieve a localization accuracy equal to the sampling rate of the $\textsc{rf}$ signals, the network is discouraged from predicting positives, which translates into a low recall (true positive rate or sensitivity).

To alleviate the stringent nature of the ground truth labels, we introduce another ground truth label, $\varphi^*$, which is generated by convolving the non-binary bubble count array $\varphi$ with a Gaussian convolution kernel $G_a$: $\varphi^* = \varphi*G_a$. Here, $G_{a,j} = e^{-a j^2}$, with $j$ the array index and $a$ a parameter that sets the kernel width (full width at half maximum $\mathrm{\textsc{FWHM}} = 2\sqrt{\mathrm{ln}(2)/a}$). We refer to the binary array $\varphi^\dagger$ as a \textit{hard label} and to the convolved bubble count array $\varphi^*$ as a \textit{soft label}~\cite{Sironi2016}. We apply the same convolution operation to the output of the network to generate soft predictions $\hat{\varphi}^*$.

The soft labels $\varphi^*$ are continuous distributions rather than binary distributions. Therefore, we use a regression loss for the soft labels. In particular, we use the $L_1$ loss, $L_1 = \frac{1}{N}\sum_{i=1}^N |\hat{\varphi}^*_i - \varphi^*_i|$, which more tightly imposes sparsity than the more common $L_2$ loss (mean squared error). 

We anticipate that the soft label approach will increase the microbubble recall. If the network predicts a bubble in the vicinity of a ground truth bubble, the soft labels $\varphi^*$ and the soft predictions $\hat{\varphi}^*$ will still overlap, which contributes to minimizing the loss. Although softening the labels comes at the expense of resolution, the sampling interval $\Delta t$ is far beyond the reach of current super-resolution approaches. It is therefore acceptable to sacrifice some resolution to increase the bubble detection rate.

\subsubsection{Dual-loss}
In summary, hard labels impose a resolution equal to the sampling rate, potentially resulting in a reduced recall. Soft labels are expected to yield a higher recall, at the expense of a reduced resolution. Therefore, we propose a dual-loss $\mathcal{L}$ that combines both label types and leverages the advantages of the two: 
\begin{equation}\label{EqTotalLoss}
\mathcal{L}= \varepsilon_1L_1(\varphi^*,\hat{\varphi}^*) + \varepsilon_2\mathrm{DL}(\varphi^\dagger,\hat{\varphi}^\dagger).
\end{equation}
where $\varepsilon_1$ and $\varepsilon_2$ are tunable proportionality constants. Fig.~\ref{blanke4} summarizes the steps involved in computing the dual loss.

\subsubsection{Training implementation}\label{SectionTrainingScheme}
We train the network on 1024 signals from the training data set (see supplementary material for the effect of training set size), using a batch size of 64. Each model is trained for 1250 epochs. We implement the network and the training in Python using PyTorch. We use the Adam optimization algorithm with a learning rate of 0.01 to optimize the network parameters. A step learning rate scheduler is used to reduce the learning rate to 0.001 for the final 250 epochs to stabilize the training. Training was executed on a 24~GB \textsc{nvidia} Quadro RTX 6000 graphics processing unit (\textsc{gpu}).

\subsection{Model evaluation}
\subsubsection{Quantitative evaluation on single \textsc{rf} signals.}\label{SectionMethodsEvaluation}
\begin{figure}[ht]
    \centering
    \includegraphics[width=0.9\columnwidth]{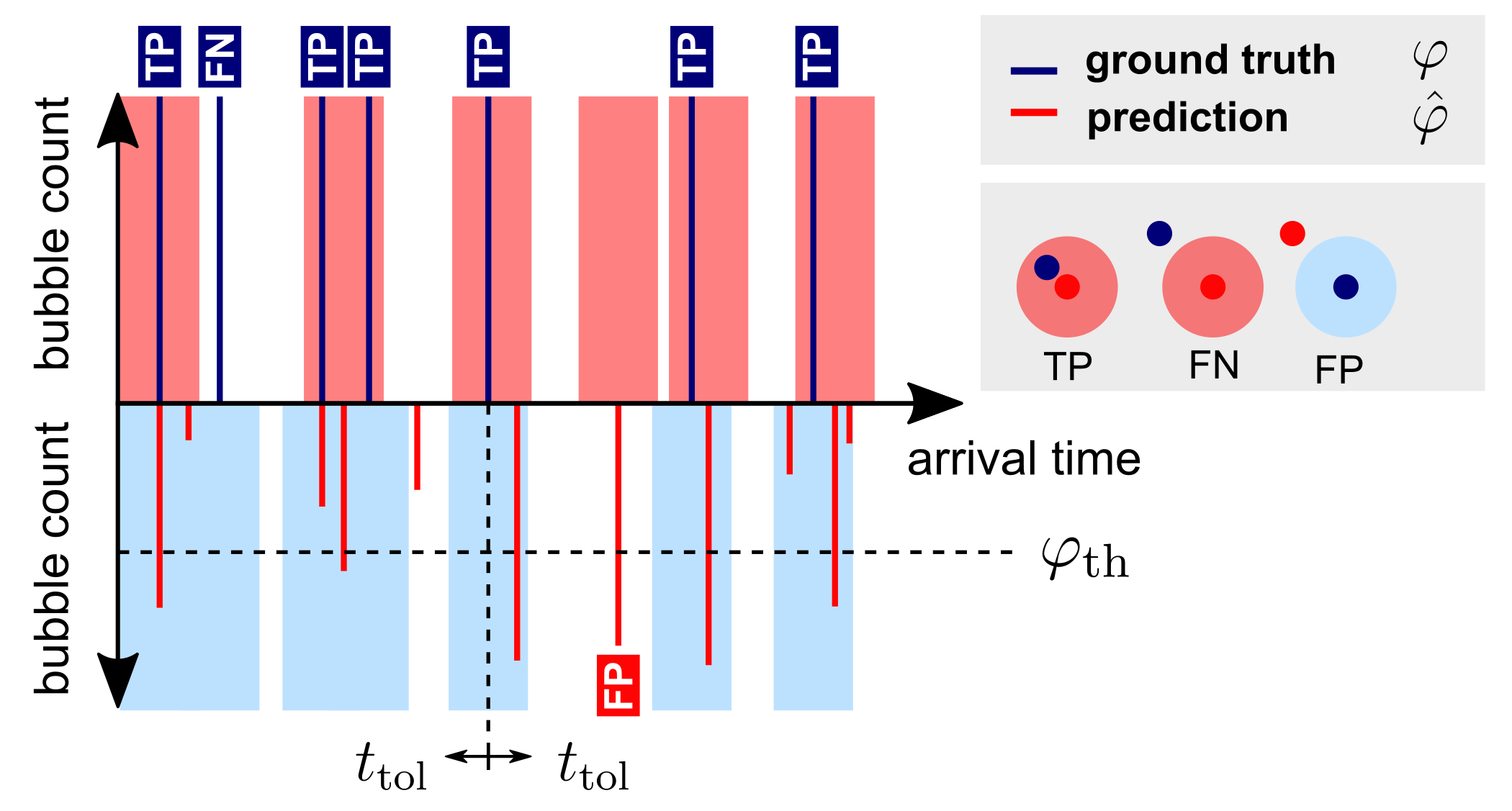}
    \caption{\textbf{Model evaluation with localization tolerance.} The detection threshold $\varphi_\mathrm{th}$ classifies predictions either as positives or negatives. The tolerance radius $t_\mathrm{tol}$ defines a tolerance region centered at a microbubble. A ground truth bubble that falls within the tolerance region of any predicted bubble counts as a true positive (TP). A ground truth bubble that does not fall within the tolerance interval of any predicted bubble counts as a false negative (FN). A predicted bubble that does not fall within the tolerance region of any ground truth bubble counts as a false positive (FP).}
    \label{blanke5}
\end{figure}

To characterize the performance of each trained neural network, we use precision, recall, and their harmonic mean, the $F_1$ score. Precision and recall are defined as follows:
\begin{equation}\label{EqRecall}
    \mathrm{Precision} = \frac{\mathrm{TP}}{\mathrm{TP} + \mathrm{FP}}
,~~
    \mathrm{Recall} = \frac{\mathrm{TP}}{\mathrm{TP} + \mathrm{FN},}
\end{equation}
where TP is the number of true positives, FP is the number of false positives, and FN is the number of false negatives. To classify grid points as positives or as negatives, we apply a detection threshold value $\varphi_{\mathrm{th}}$ to the continuous predictions $\hat{\varphi}$ (Fig.~\ref{blanke5}). The $F_1$ score is the harmonic mean of precision and recall:
\begin{equation}
    F_1 = \frac{2 \times \mathrm{Precision} \times \mathrm{Recall}}{\mathrm{Precision} + \mathrm{Recall}} = \frac{2\mathrm{TP}}{2\mathrm{TP} + \mathrm{FP} + \mathrm{FN}}
\end{equation}
The $F_1$ score (and precision and recall) depends on the choice of $\varphi_{\mathrm{th}}$. As the optimal $\varphi_{\mathrm{th}}$ differs between models, a fixed $\varphi_\mathrm{th}$ could introduce bias in comparisons between models. Therefore, we make $\varphi_\mathrm{th}$ a hyperparameter that is optimized for each network to maximize the average $F_1$ score on the training set. Next, $F_1$ is computed on the test set with this threshold. 

Since it is favorable to accept a tolerance in the bubble localization (Section~\ref{SectionSoftLoss}), we incorporate a tolerance radius $t_\mathrm{tol}$ into the definitions of true positive, false positive, and false negative (Fig.~\ref{blanke5}). A true positive is any ground truth bubble that is located within $t_\mathrm{tol}$ from a predicted bubble. A false positive is any predicted bubble that is not located within $t_\mathrm{tol}$ from a ground truth bubble. A false negative is any ground truth bubble that is not located within $t_\mathrm{tol}$ from a predicted bubble. For fair comparison, we optimize $\varphi_{\mathrm{th}}$ for each $t_\mathrm{tol}$. 

\subsubsection{Image reconstruction}\label{SectionMethodsReconstruction}

To assess the impact of our 1D super-resolution approach to ultrasound imaging, we generate 2D channel data by extending the simulator to output all receive channels of the virtual transducer. We apply a trained model to each \textsc{rf} signal independently and, subsequently, apply a basic delay-and-sum beamforming (\textsc{das}) beamforming algorithm to the deconvolved \textsc{rf} signals to reconstruct an ultrasound image.

\section{Results}

\subsection{Dual loss and model performance}\label{ResultsDualLoss}

\subsubsection{Evaluated models}
We present a quantitative comparison of seven different models that share architecture (Section~\ref{SectionMethodsArchitecture}) but were trained with different loss functions. More precisely, we varied the loss coefficients $\varepsilon_1$ and $\varepsilon_2$ in Eq.~\ref{EqTotalLoss}, and the Gaussian width parameter $a$ (Section~\ref{SectionDualLoss}). Setting $\varepsilon_1 = 0$ results in pure hard label training (Dice loss), whereas setting $\varepsilon_2 = 0$ results in pure soft label training ($L_1$~loss). For dual-loss training, $\varepsilon_1 = 1$ and $\varepsilon_2 = 1.6$. We empirically determined that this ratio of $\varepsilon_1$ and $\varepsilon_2$ leads to fast convergence. Table~\ref{TableModelOverview} details the training configuration for each model.

 \begin{table}[t]
 \caption{Properties of the evaluated models} 
\label{TableModelOverview}
\centering
\begin{tabular}{@{\extracolsep{4pt}}lcclrr}
\toprule   
{\textbf{ID}} & \multicolumn{2}{c}{\textbf{Loss function}}  & \multicolumn{3}{c}{\textbf{Soft kernel properties}}\\
 \cmidrule{2-3} 
 \cmidrule{4-6} 
 {} & Hard & Soft  & $a$ & \textsc{fwhm} & \textsc{fwhm} \\ 
 {} & labels & labels & {} & (pts.) & {/$t_\lambda$} \\
\midrule
H           & \checkmark & -          & -    & - & - \\ 
\cmidrule{1-6}
S$_{0.01}$  & -          & \checkmark & 0.01 & 16.7 & 0.226 \\ 
S$_{0.1}$   & -          & \checkmark & 0.1  & 5.3  & 0.072  \\ 
S$_{1}$     & -          & \checkmark & 1    & 1.7  & 0.023  \\ 
\cmidrule{1-6}
HS$_{0.01}$ & \checkmark & \checkmark & 0.01 & 16.7 & 0.226 \\ 
HS$_{0.1}$  & \checkmark & \checkmark & 0.1  & 5.3  & 0.072 \\ 
HS$_{1}$    & \checkmark & \checkmark & 1    & 1.7  & 0.023 \\ 
\bottomrule\\
\end{tabular}

\begin{tabular}{@{\extracolsep{4pt}}lcclrl}
\multicolumn{6}{p{240pt}}{The model identifiers (ID) indicate whether the model was trained with hard labels (H), soft labels (S), or with both (HS). The subscripts indicate the width parameters $a$ of the Gaussian soft kernels. The full width at half maximum (\textsc{fwhm}) of these kernels is given in both the number of grid points and as a fraction of the diffraction time $t_\lambda$}
\end{tabular}

\end{table}

\begin{figure*}[ht]
    \centering
    \includegraphics[width=\textwidth]{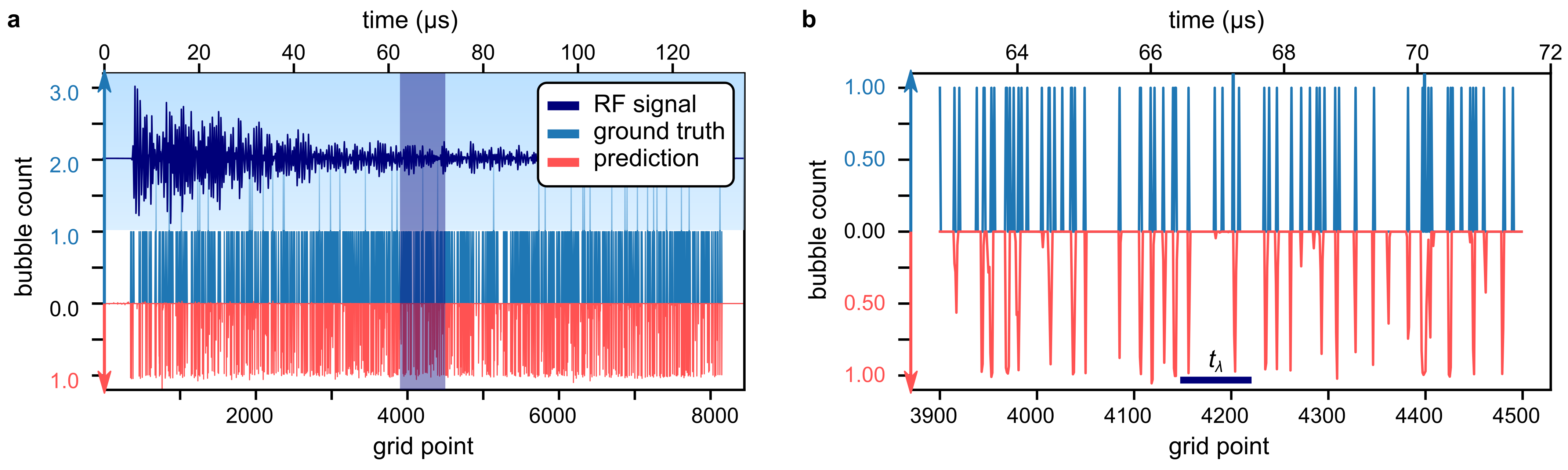}
    \caption{\textbf{Example of network output.} Model HS$_{0.1}$ applied to an \textsc{rf} signal from the test set. \textbf{a} Full-length \textsc{rf} signal $V$, ground truth $\varphi$, and prediction $\hat{\varphi}$. The horizontal axis is shared. The ground truth contains 910 microbubbles. \textbf{b} Enlargement of the shaded area in \textbf{a}. The scale bar indicates $t_\lambda$ (two ultrasound cycles).}
    \label{blanke6}
\end{figure*}

\subsubsection{Precision-recall trade-off and detection-localization trade-off}
\begin{figure*}[ht]
    \centering
    \includegraphics[width=\textwidth]{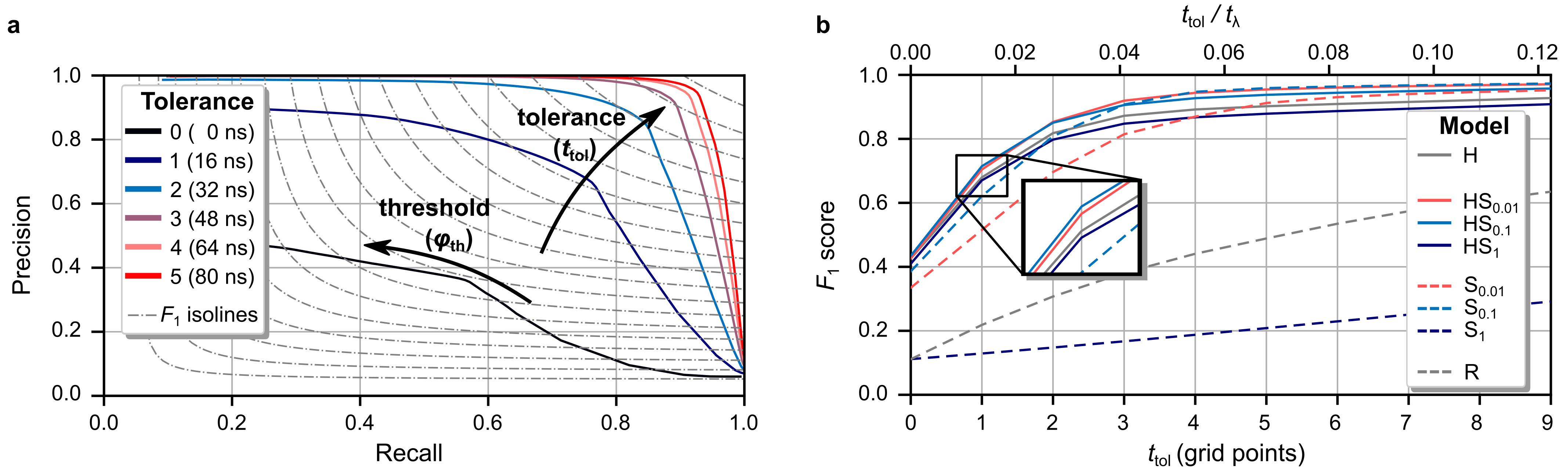}
    \caption{\textbf{Model evaluation on test set.} \textbf{a} Average precision-recall curves of model HS$_{0.1}$. The dashed lines are lines of constant $F_1$ score. \textbf{b} Comparison of the $F_1$-$t_\mathrm{tol}$ curves of the different models. The model identifiers are explained in Table~\ref{TableModelOverview}. Model R is a reference model, which outputs an array of random numbers between 0 and 1.}
    \label{blanke7}
\end{figure*}

Fig.~\ref{blanke6} shows a typical example of an input \textsc{rf} signal, its ground truth bubble distribution $\varphi$, and the output $\hat{\varphi}$ of a trained model (HS$_{0.1}$) for this signal. The figure shows that the network successfully detects bubbles separated by far less than a wavelength and that it does so with a high localization accuracy. To demonstrate the dependence of precision and recall on $\varphi_\mathrm{th}$ and $t_\mathrm{tol}$ we apply model HS$_{0.1}$ to the test set (960 signals). For each prediction, we compute precision and recall and average these over the test set. We repeat this for a range of detection thresholds $\varphi_\mathrm{th}$. Fig.~\ref{blanke7}a shows the resulting precision-recall curves. As $\varphi_\mathrm{th}$ increases, precision increases, but recall decreases. 

In addition to the trade-off between precision and recall, there also exists a trade-off between detection and localization: precision, recall, and $F_1$ score increase with increasing $t_\mathrm{tol}$. Super-resolution imaging with bubbles involves two tasks: detection \textit{and} localization. Super-resolution performance can only be expressed in terms of both. Fig.~\ref{blanke7}a shows that the precision-recall curves rapidly converge for increasing $t_\mathrm{tol}$: a precision and recall both as high as 0.90 can be achieved with a localization tolerance of only 3 grid points, corresponding to $t_\mathrm{tol}/t_\lambda = 0.04$.

\subsubsection{Model comparison}
To compare the different models summarized in Table~\ref{TableModelOverview}, we plot the $F_1$-$t_\mathrm{tol}$~curve of each model (Fig.~\ref{blanke7}b). A random model R serves as a baseline reference. This model simply outputs an array of random numbers between 0 and 1. For each model and tolerance, we set $\varphi_\mathrm{th}$ by maximizing the average $F_1$ on the training set.

Two models stand out: HS$_{0.01}$ and HS$_{0.1}$. They yield the best average $F_1$ score over the investigated range of $t_\mathrm{tol}$. Model H performs well for small $t_\mathrm{tol}$ but underperforms for large $t_\mathrm{tol}$. This reflects the absent localization tolerance of the hard labels. Conversely, models S$_{0.01}$ and S$_{0.1}$ perform well for large $t_\mathrm{tol}$, but underperform for small $t_\mathrm{tol}$. The absence of hard labels during training lowers the localization accuracy of these models. Model HS$_1$ performs worse than model H, even though it was trained with the dual loss. A possible explanation is that the soft kernel is too narrow to be effective. Model S$_1$ failed to learn its task, predicting a constant value for all but a few grid points, which explains the low score. An explanation for this failed training is that the signal was too sparse (mostly zero due to the narrow soft-label kernel) to use a regression loss. 

As hypothesized, the proposed dual loss benefits from the advantages of both hard labels and soft labels. It is effective if the soft convolution kernel used for training is sufficiently wide. Since model HS$_{0.01}$ has a slightly higher $F_1$ score than HS$_{0.1}$ (averaged over the investigated $t_\mathrm{tol}$), we select HS$_{0.01}$ for further analysis.

\begin{figure*}[ht]
    \centering
    \includegraphics[width=\textwidth]{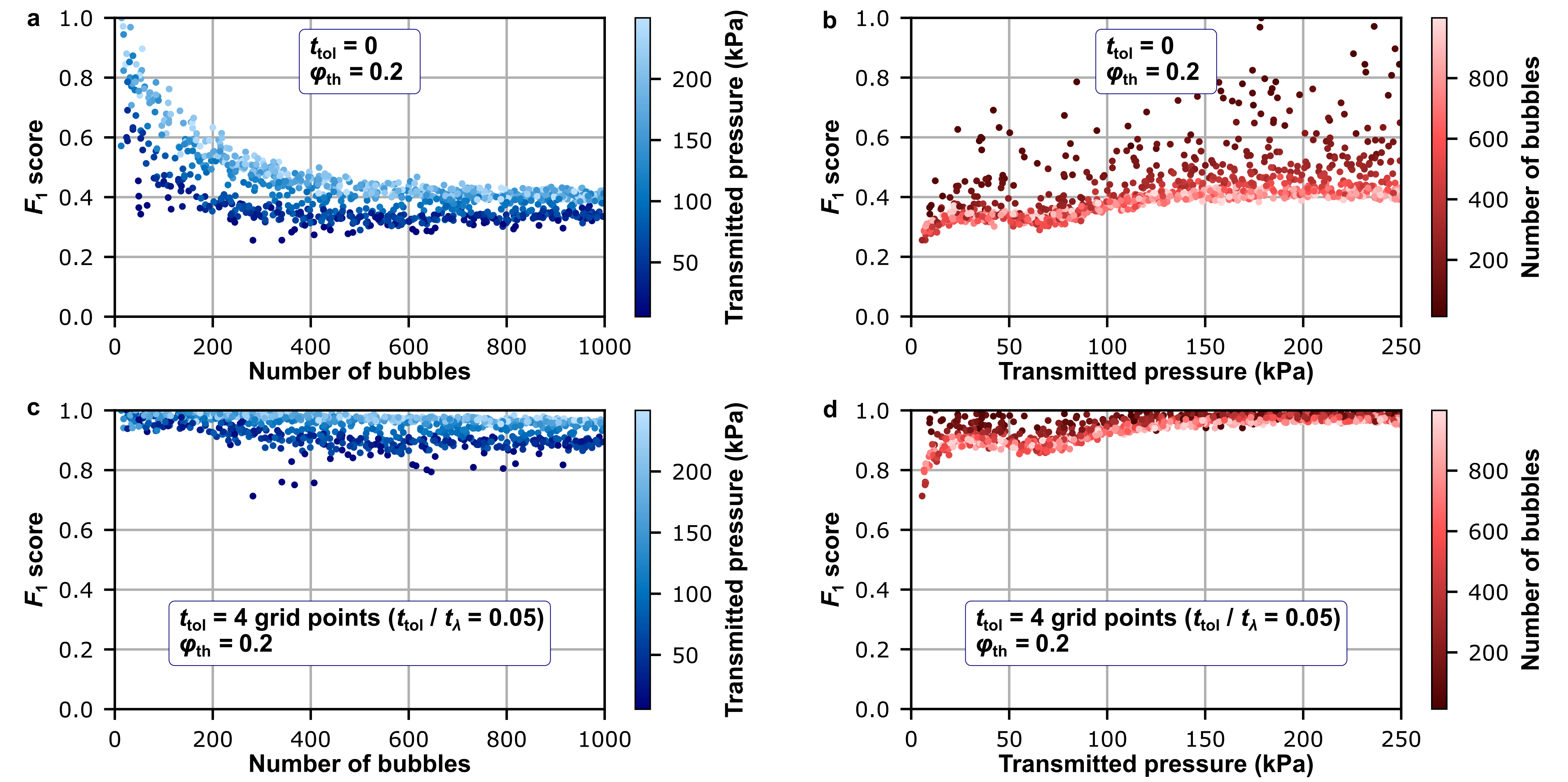}
    \caption{\textbf{Network performance as a function of acoustic pressure and number of microbubbles.} Performance of model HS$_{0.01}$ applied to individual signals from the test set for $t_\mathrm{tol}$ = 0 (\textbf{a} and \textbf{b}) and for $t_\mathrm{tol}$ = 4 grid points (\textbf{c} and \textbf{d}).}
    \label{blanke8}
\end{figure*}

\begin{figure}[htb]
    \centering
    \includegraphics[width=\columnwidth]{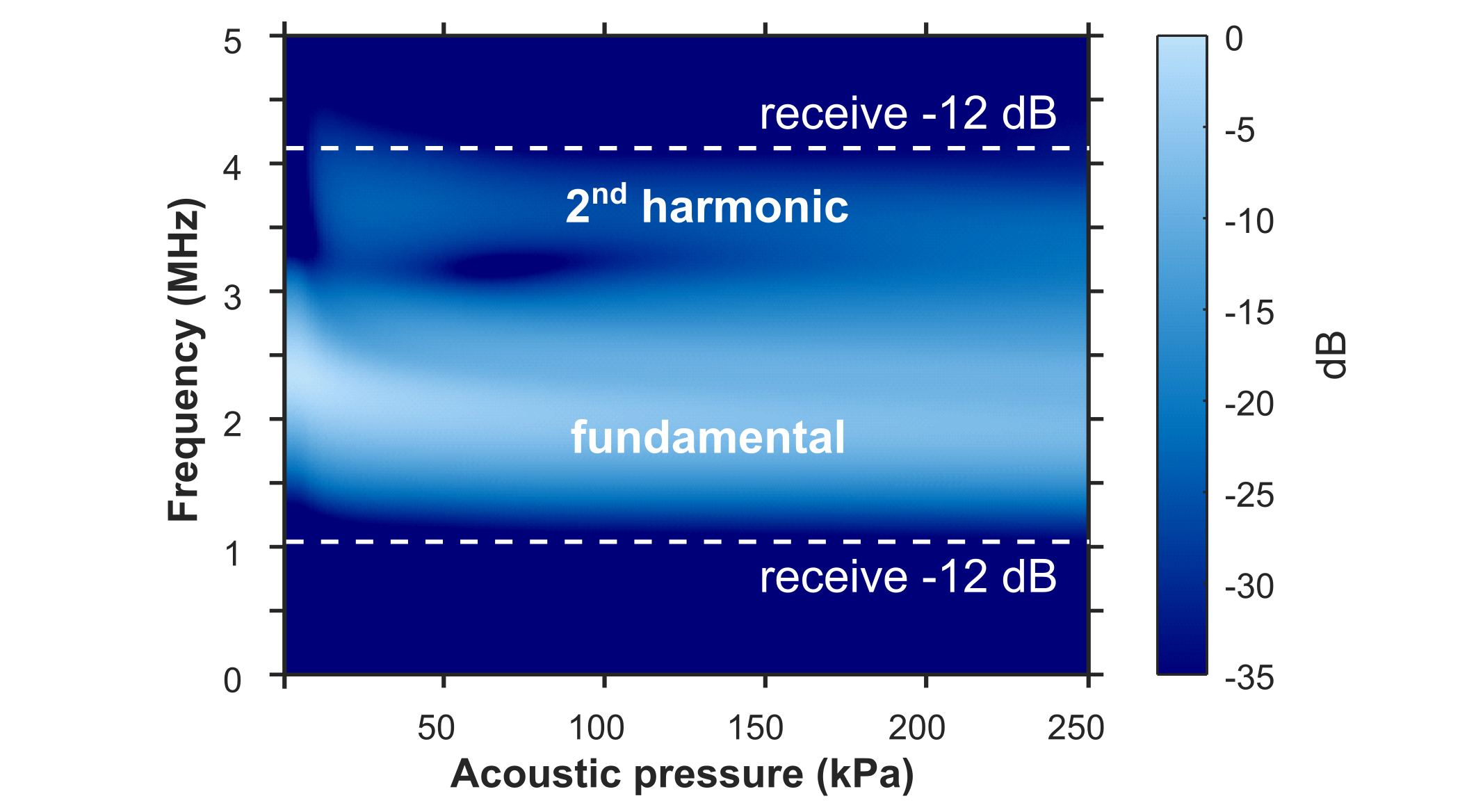}
    \caption{\textbf{Frequency spectrum of a single microbubble echo.} Simulated frequency spectrum (absolute value of the Fourier transform) of the bubble response as a function of acoustic pressure. The bubble has a radius of 2.4~$\mu$m and is placed 3.7~mm from the transducer. Each echo is normalized by the acoustic transmit pressure, and the full spectrum is normalized by its maximum value. The dashed lines indicate the -12~dB points of the receive transfer function of the transducer. Note that only the fundamental and second harmonic responses fall within this region.}
    \label{blanke9}
\end{figure}

\subsection{Effect of microbubble density and acoustic pressure}

The simulated data sets cover a large range of bubble concentrations (10-1000 microbubbles per \textsc{rf} signal) and acoustic pressures (5-250 kPa). To investigate the model performance as a function of these two parameters, we apply model HS$_{0.01}$ to the test set and compute the $F_1$ score for each \textsc{rf} signal individually.  Fig.~\ref{blanke8} displays the $F_1$ score of each prediction from the test set as a function of the number of bubbles in the ground truth $\varphi$ and the transmitted acoustic pressure for two selected localization tolerances. Here, we discuss two striking observations.

The first observation is that, for $t_\mathrm{tol} = 0$, the $F_1$ score depends strongly on the bubble concentration. The $F_1$ score is close to 1 for low bubble concentrations, and it steeply drops to about 0.4 for higher bubble concentrations where it reaches a plateau. With $t_\mathrm{tol} = 4$ grid points ($t_\mathrm{tol}/t_\lambda=0.05$), the $F_1$ score still decreases with bubble concentration but now exceeds 0.8 for all bubble concentrations. Therefore, a larger $t_\mathrm{tol}$ is necessary to achieve an adequate ($F_1>0.9$) detection at higher concentrations. In other words, the super-resolution performance of the network is lower for high concentrations. The strong dependence of the $F_1$ score on bubble concentration can be interpreted in terms of overlapping point spread functions. As discussed in Section~\ref{SectionGeometry}, $N_\mathrm{MB}=10$ yields a low probability that any two microbubble echoes overlap. Conversely, with $N_\mathrm{MB}=1000$, the average overlap between two adjacent bubble echoes is 94\%, complicating the detection and localization task.

The second observation from Fig.~\ref{blanke8} is that the model performs better for high acoustic pressures. We hypothesize that the reduced network performance for low acoustic pressure is related to nonlinear microbubble behavior. The frequency spectrum of a single microbubble (Fig.~\ref{blanke9}) shows two major differences between low ($<10~$kPa) and high ($>10~$kPa) acoustic pressure amplitudes $P_\mathrm{ac,max}$ that are due to nonlinearities and that could both explain the reduced network performance. Firstly, the center frequency of the fundamental response is higher for low $P_\mathrm{ac,max}$. Pressures between 5 and 10~kPa represent only 2\% of the training signals, which could have led to data imbalance during training. Secondly, the second harmonic is not observed for low $P_\mathrm{ac,max}$. As higher harmonics are sensitive to the acoustic pressure, they could help the network to estimate the acoustic pressure implicitly and thus the local convolution kernel. Although the exact origin of this behavior requires further investigation, we note that the clinical relevance of pressures below 10~kPa is limited. A second increase in network performance is observed between 70 and 100~kPa. This may be related to the fact that the harmonic content appears to be independent of acoustic pressure above 100~kPa, but this behavior also needs further investigation.

\begin{figure*}[ht]
    \centering
    \includegraphics[width=\textwidth]{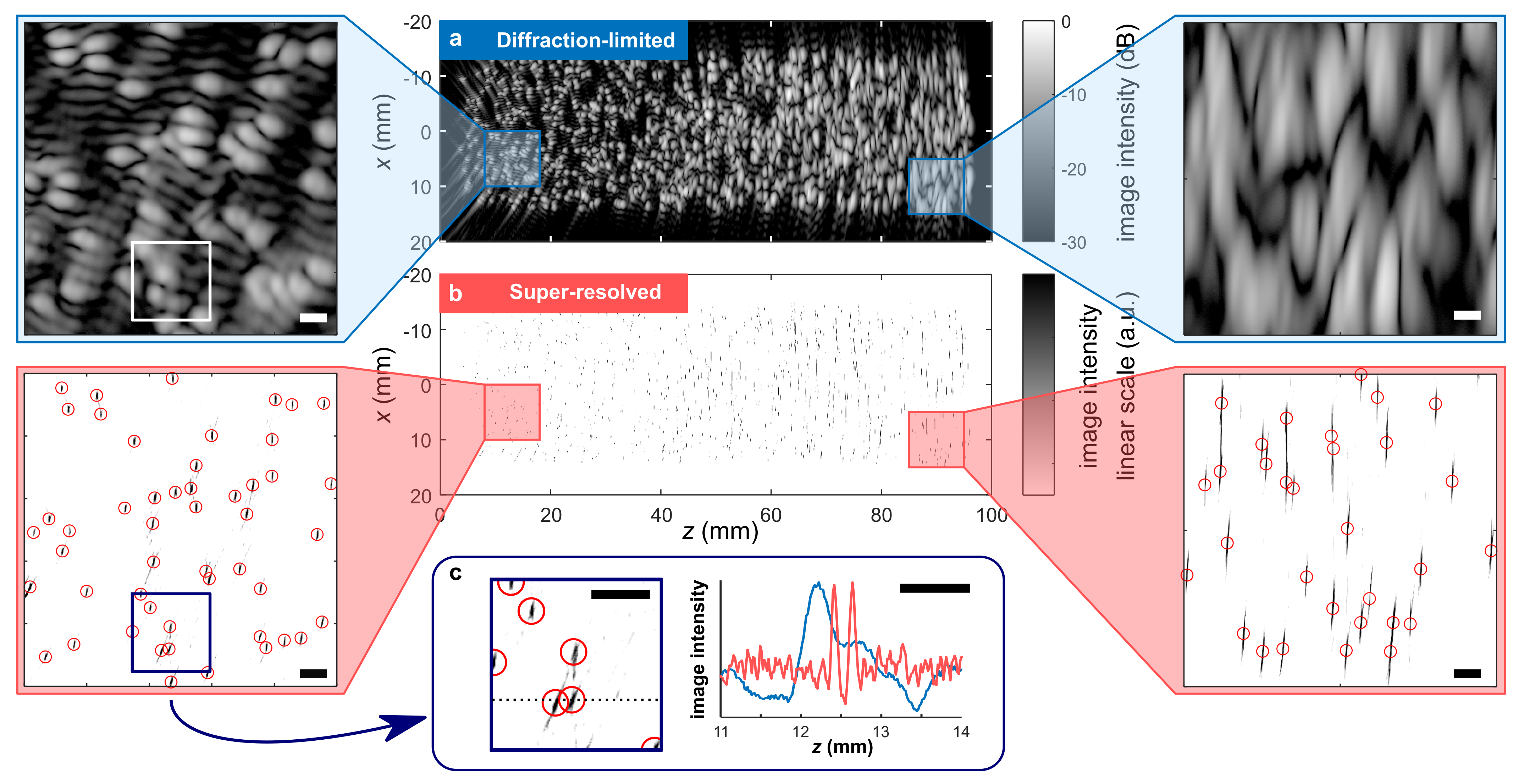}
    \caption{\textbf{Delay-and-sum reconstruction of original (a) and deconvolved (b) channel data.} The ground truth bubble locations are indicated with red circles. The bubble cloud contains 907 bubbles. The transmit pressure was 205~kPa. The white scale bars indicate one wavelength of the transmit pulse. The color scale in \textbf{b} is inverted to improve the visibility of the small dots (black on white). \textbf{c} Intensity profile of indicated cross-section.}
    \label{blanke10}
\end{figure*}

\subsection{Super-resolved delay-and-sum reconstruction} 

In this section, we demonstrate how deconvolved \textsc{rf} signals can be combined to generate a super-resolved ultrasound image using a basic delay-and-sum beamforming algorithm, as described in Section~\ref{SectionMethodsReconstruction}. We apply model HS$_{0.01}$ to each \textsc{rf} signal independently and, subsequently, apply \textsc{das} to the deconvolved \textsc{rf} signals. For comparison, we also apply the delay-and-sum to the original element data (i.e. without applying the model).  

DAS beamforming with the original element data produces a standard ultrasound B-mode speckle image, consisting of overlapping point spread functions whose widths increase with depth (Fig.~\ref{blanke10}a). The super-resolved image (Fig.~\ref{blanke10}b) demonstrates the gain in localization accuracy achieved with our method. The magnified regions, near and far from the transducer, show the excellent agreement between the super-resolved image and the ground truth.   Fig.~\ref{blanke10}c zooms further on the image and shows two clearly distinguishable microbubbles. The full width at half maximum (\textsc{fwhm}) of the left point spread function is 95~µm, which is 11\% of the wavelength and an order of magnitude smaller than the \textsc{fwhm} of the diffraction-limited point spread function. Also, note that the maxima in the diffraction-limited intensity cross-section (from the B-mode image) do not coincide with the actual microbubble locations due to interference. The microbubbles are accurately localized in the super-resolved image.

The axial resolution of the super-resolved image is excellent throughout the bubble cloud. By contrast, the lateral resolution deteriorates with increasing depth, as it also does for the diffraction-limited image. Near the transducer, where the $f$-number (depth divided by aperture) is low, the localization uncertainty in the time domain translates to a low uncertainty in the lateral direction ($\textsc{fwhm}\sim170$~µm). Far from the transducer, where the $f$-number is high, the same localization uncertainty in the time domain translates to high uncertainty in the lateral direction ($\textsc{fwhm}\sim1.2$~mm). 

Finally, the detection rate of the microbubble detection after reconstruction is, visually, higher than in a single \textsc{rf} signal (e.g. Fig.~\ref{blanke6}). The delay-and-sum beamforming combines estimated bubble distributions from multiple transducer channels such that the detection uncertainty decreases. 

\subsection{The effects of signal noise and polydispersity}

\begin{figure}[ht]
    \centering
    \includegraphics[width=0.95\columnwidth]{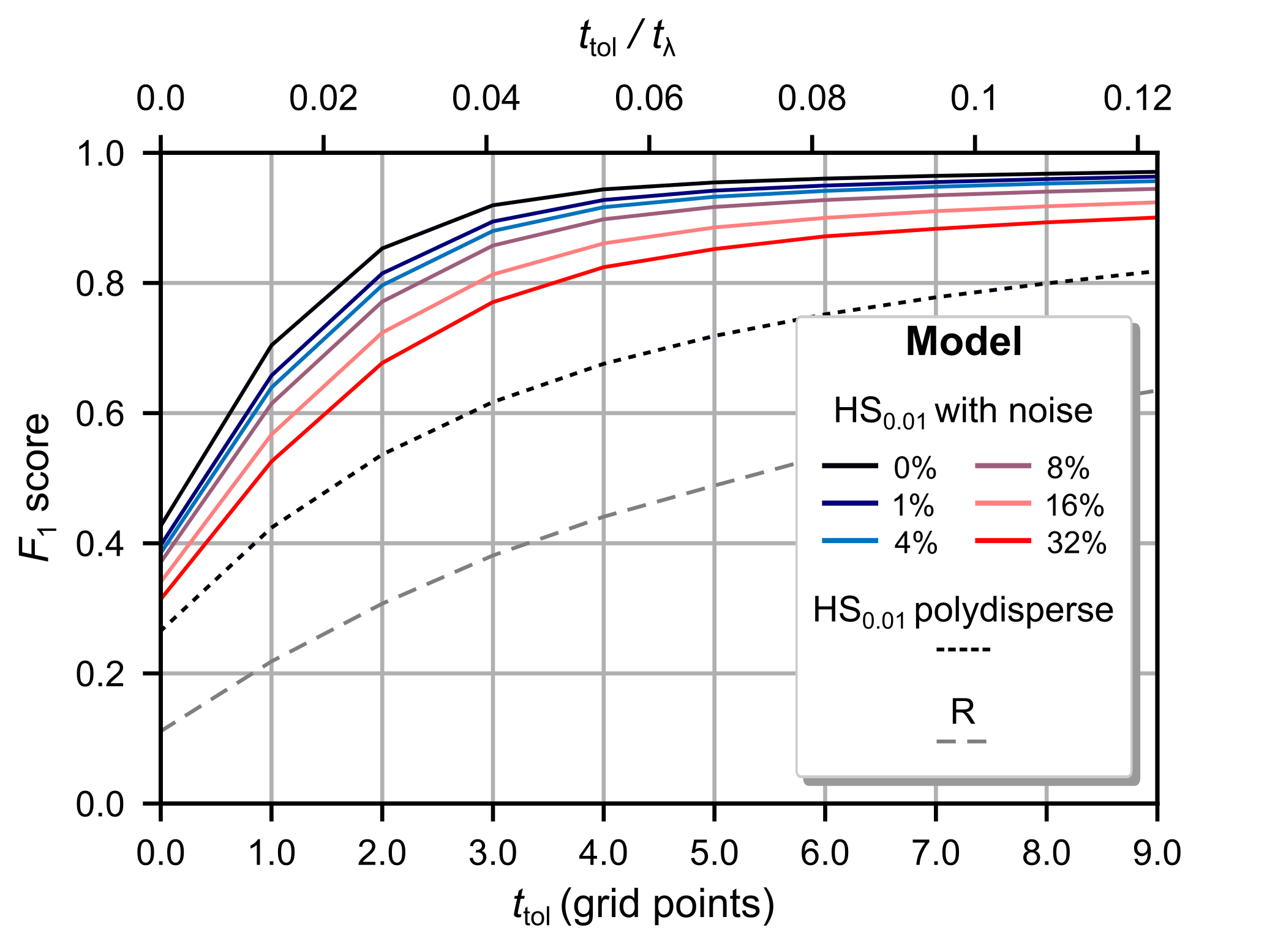}
    \caption{\textbf{The effects of noise and polydispersity.} Comparison of the $F_1$-$t_\mathrm{tol}$ curves of model HS$_{0.01}$ trained and tested on different data sets. The $F_1$-$t_\mathrm{tol}$ curves are evaluated on the test set. The solid lines indicate models trained and tested on data from monodisperse microbubbles but with increasing levels of noise. The percentages indicate the \textsc{rms} noise levels with respect to $U_\mathrm{ref}$.  The dotted line indicates the model trained and tested on data from polydisperse microbubbles. Model R is the same random reference model as in Fig.~\ref{blanke7}b.}
    \label{blanke11}
\end{figure}

To assess the applicability of our method to experimental data, we investigate the robustness of model HS$_{0.01}$ to added signal noise. To this end, the model is retrained on the original \textsc{rf} data, to which noise is added ad hoc, i.e. the noise is different for each training epoch. First, Gaussian white noise with an \textsc{rms} value $U_\mathrm{noise}$ is added to the \textsc{rf} signals. Next, a 4th degree Butterworth low-pass filter with a cutoff frequency of 5.1~MHz filters the signals. Since the clean signals are bandlimited, the low-pass filter only affects the noise. After training the network, the threshold $\varphi_\mathrm{th}$ is optimized on the training set, and the network is tested on the test set (both with added noise). This procedure of training, threshold optimization, and testing is repeated for different values of $U_\mathrm{noise}$. Fig.~\ref{blanke11} shows the $F_1$-$t_\mathrm{tol}$ curves evaluated on the test set for different noise levels. The \textsc{rms} value of the Gaussian white noise, $U_\mathrm{noise}$, is expressed as a percentage of a reference signal value $U_\mathrm{ref}$. This reference value is the maximum signal value recorded from a single microbubble of $R_0 = 2.4$~µm at 5~cm from the transducer that is driven at 125~kPa, which is a typical signal value in the data set. 

To estimate typical experimental values of $U_\mathrm{noise}$ and $U_\mathrm{ref}$, we performed a simple experiment with a dilute solution ($\sim 10$~microbubbles/mL) of $R_0 = 2.4$~µm monodisperse bubbles (DSPC:DPPE-PEG5000, 9:1). We acquired data with a P4-1 transducer (Philips ATL) connected to a Verasonics Vantage 256 system. We achieved the best signal-to-noise ratio with the \textsc{tgc} (time-gain compensation) values set to maximum. Based on these experiments, we estimate that $U_\mathrm{noise}$ is 15\% of $U_\mathrm{ref}$. Comparing this value to the results in Fig.~\ref{blanke11}, we expect that our method can be applied to experimental data, provided the \textsc{rf} signal simulator represents the experimental data accurately enough.

Another question relevant to the practical application of our method is how well the network performs on \textsc{rf} signals obtained from a polydisperse population of microbubbles. To answer this question, we generated a new dataset. We modified the simulator to sample the microbubble radii from the size distribution of BR-14 bubbles~\cite{Segers2018b}. We found that this distribution is described well by the curve $AR_0^2e^{-bR_0}$, with $b=2.19$~µm$^{-1}$ and $A$ a normalization factor. We excluded bubbles with a radius less than 0.5~µm because they generate signals more than 40~dB weaker than 2.4~µm bubbles and require unacceptably long simulation times. We generated new data sets for training, validation, and testing. We retrained and tested model HS$_{0.01}$ following the same procedure as before. The test results are shown in Fig.~\ref{blanke11}. The network performance on data from monodisperse bubbles is better than on data from polydisperse bubbles. We attribute this better performance to a lower degree of kernel variability: the individual bubble echoes vary less in amplitude and shape.


\section{Discussion}

We have presented a dilated convolutional neural network that can be trained for super-resolved detection and localization of microbubbles in individual simulated \textsc{rf} channels with up to 1000 bubbles per signal, a density that corresponds to an average overlap of 94\% between two adjacent bubble echoes.

In expressing the super-resolution performance, there exists a trade-off between localization accuracy and detection: a larger localization tolerance yields a higher detection rate. Nonetheless, this trade-off is by no means restrictive with respect to super-resolution: a localization tolerance corresponding to 7\% of the wavelength already yields an average $F_1$ score of 0.95. The potential of our approach is demonstrated by a delay-and-sum reconstruction with deconvolved \textsc{rf} signals: the resulting image shows order-of-magnitude gain in axial resolution compared to a standard B-mode image. 

In this work, we decomposed super-resolution ultrasound imaging into two consecutive steps: deconvolving \textsc{rf} data with a neural network and beamforming. The first step was the main focus. For beamforming, we used a standard delay-and-sum reconstruction. We demonstrated that the correlations between multiple single-channel deconvolved \textsc{rf} signals reduced the uncertainty in microbubble detection and localization. Therefore, a logical extension would be to use a neural network for beamforming as well. In fact, the whole pipeline from raw data to ultrasound image could be performed by a neural network. We expect that a neural network trained with 2D element data can image even higher bubble concentrations. For example, within the presented geometry, $N_\mathrm{MB}=1000$ corresponds to a bubble density of 0.3/$\lambda^2$ in space, while it corresponds to an echo density of $9/t_\lambda$ in a single-channel \textsc{rf} signal. Also, the lateral imaging resolution in the current two-step approach still depends on the beamforming algorithm. Therefore, we expect that a 2D end-to-end approach can improve the lateral imaging resolution as well. We also expect that a 2D end-to-end approach would be more robust to noise as the network could filter out uncorrelated noise between single-element signals. Furthermore, we envision that our super-resolution approach can be interfaced with existing \textsc{ulm} techniques to achieve an even higher gain in resolution.

Applying a neural network to 2D element data poses computational challenges. For example, Youn \textit{et al.} have already made efforts to achieve super-resolved images directly from the element data. They simulated data from linear scatterers~\cite{Youn2020} acquired with three acquisitions. They studied similar scatterer densities per square wavelength (up to 0.4/$\lambda^2$), but we simulated a larger domain with a higher sampling rate, resulting in much larger 2D element data ($8446\times96$ compared to $265\times64\times3$). With present computational resources, this large data size puts strict limits on neural network and batch size. 

Further refinements to the simulations and the beamforming algorithm and further system characterization are necessary to apply our method successfully to experimental data. For example, 3D wave propagation simulations are needed to accurately describe the local variations in acoustic pressure. Acoustic pressure inhomogeneity is especially pronounced in strongly attenuating tissue. Furthermore, the transmit pulse may need to be characterized more precisely to simulate more realistic training data. Accurate system characterization is also needed to compensate for changes in receive gain.

Evaluating experimental results is challenging because obtaining an experimental ground truth is near-impossible. A ground truth could be obtained by confining the microbubbles, which would change their response. The models describing microbubble dynamics presented here are only valid for bubbles in the free field. Accurate models to describe the dynamics of confined microbubbles are not readily available.

An important question that remains to be answered is how well the neural network can exploit the nonlinear bubble echoes to differentiate the microbubbles from a background of linear scatterers. To differentiate microbubbles (flowing) from the background (static), \textsc{ulm} often relies on spatio-temporal filtering using an image sequence. By contrast, the harmonics in the microbubble response allow for differentiating them from the background within a single frame, which would provide a major step towards real-time super-resolution imaging. To answer this question, quantitative characterization of scatterer strength in various tissues is required to generate a realistic training data set.

It takes the trained neural network 0.08 seconds to deconvolve all 96 single-element \textsc{rf} signals used for Fig.~\ref{blanke10}b (in 2 batches of 48 signals on the \textsc{nvidia} Quadro RTX 6000). For the subsequent \textsc{das} reconstruction, we used a suboptimal routine. In 2014, optimized beamforming algorithms running on \textsc{gpu}s could achieve frame rates well over 1000 frames per second~\cite{Tanter2014}. However, a smaller pixel size is needed to accommodate super-resolved imaging. For example, Fig.~\ref{blanke10}b is 2413x5744 pixels, two orders of magnitude more than a typical ultrasound image. Therefore, we estimate that the beamforming process of such an image with \textsc{gpu} computations takes on the order of 0.1~s. Considering the steadily increasing performance of \textsc{gpu}s, we believe real-time super-resolved imaging is within reach, which could reveal detailed flow structure in larger vessels in deep tissue.


\section{Conclusion}

The presented results demonstrate that the application of a convolutional neural network to transducer element data is a promising path towards super-resolved imaging of high-density microbubble populations. This work provides a concept for real-time, super-resolved ultrasound imaging with a resolution of 10 times ($F_1=0.97$) down to 74 times ($F_1=0.43$) smaller than the imaging wavelength, depending on the desired detection rate and precision.

\appendices

\section*{Acknowledgment}
We thank Chris de Korte and Tim Segers for constructive discussions. We thank Alina Kuliesh for valuable feedback on the ultrasound simulator.

\bibliographystyle{ieeetr}
\bibliography{References.bib}

\end{document}


\section*{Super-Resolved Microbubble Localization in Single-Channel Ultrasound RF Signals Using Deep Learning -- Supplementary Text}

\noindent \textbf{Nathan Blanken, Jelmer M. Wolterink, Hervé Delingette, Christoph Brune, Michel Versluis, and Guillaume Lajoinie}

\subsection{Transfer functions}

\begin{figure*}[ht]
    \centering
    \includegraphics[width=0.8\textwidth]{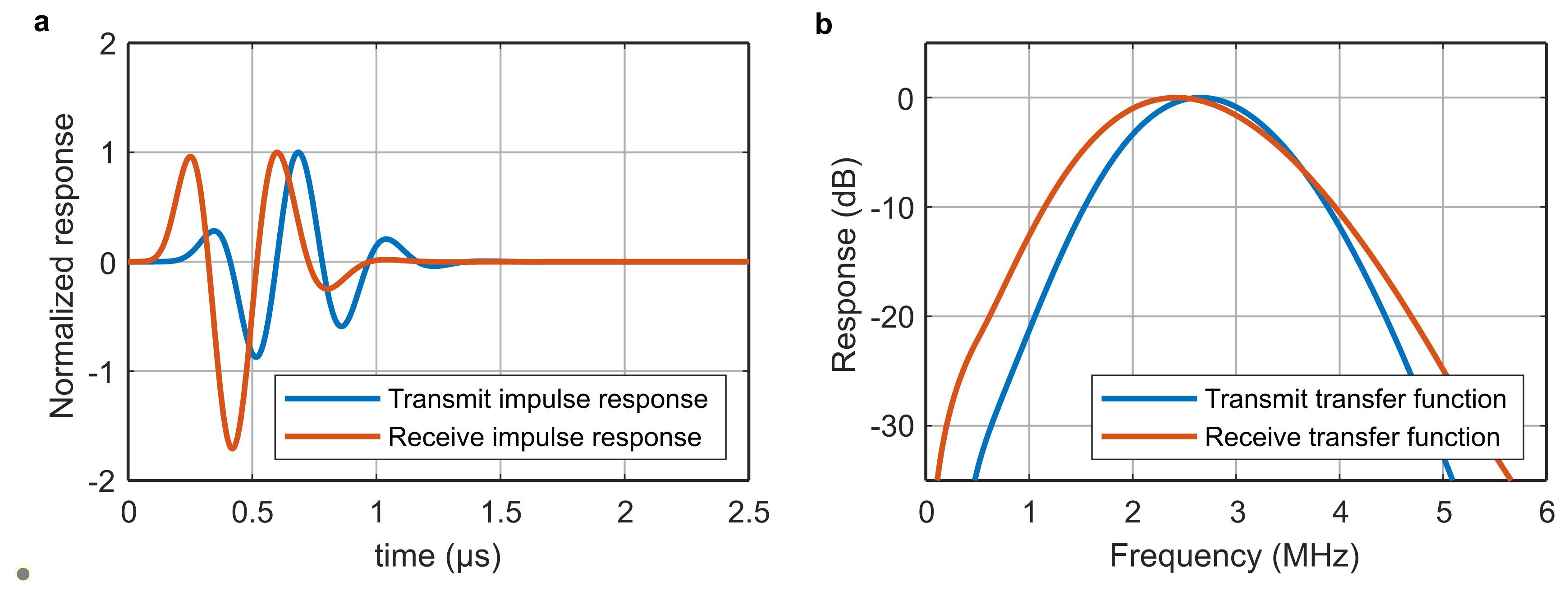}
    \caption{\textbf{Transmit and receive characteristics of the virtual transducer.} \textbf{a} Transmit and receive impulse responses. \textbf{b} Transmit and receive amplitude transfer functions.}
    \label{blankeS1}
\end{figure*}

We model the transmit and the receive characteristics of the virtual transducer upon the characteristics of a P4-1 transducer (Philips ATL). The transmit and receive impulse responses used for the \textsc{rf} data simulations are displayed in Fig.~\ref{blankeS1}a. Taking the Fourier transform of an impulse response gives the transfer function. The amplitude transfer functions are shown in Fig.~\ref{blankeS1}b.

\subsection{Second-harmonic generation}

For a medium with attenuation $\alpha(\omega) = \alpha_0\omega^2$, with $\omega$ the angular frequency, the Burgers equation can be written as~\cite{Blackstock1985,Szabo2004}:
\begin{equation}
    \frac{\partial p}{\partial z} = \frac{\beta}{\rho_\mathrm{L}c^3}p\frac{\partial p}{\partial \tau} + \alpha_0 \frac{\partial^2 p}{\partial \tau^2},
\end{equation}
where $p(z,\tau)$ is the acoustic pressure, $z$ is the coordinate along the propagation direction, $\tau$ is the retarded time, $\rho_\mathrm{L}$ is the density of the medium, and $c$ is the speed of sound.
Let $p_0(t)$ be the pressure at the transducer surface. To solve for $p(z,\tau)$, we use a perturbation approach similar to~\cite{Blackstock1985}: $p = p_1 + p_2 + ... = \epsilon\rho_\mathrm{L} c^2 + \epsilon^2\rho_\mathrm{L} c^2 + ...$, with $\epsilon << 1$ . To first order:
\begin{equation}
    \frac{\partial p_1}{\partial z} = \alpha_0 \frac{\partial^2 p_1}{\partial \tau^2}.
\end{equation}
which has the solution 
\begin{equation}
p_1(z,\tau) = \mathcal{F}^{-1}\left\{\mathcal{F}\{p_0(\tau)\}e^{-\alpha_0\omega^2z}\right\}.
\end{equation}
where $\mathcal{F}$ and $\mathcal{F}^{-1}$ denote the temporal Fourier transform and its inverse, respectively. Provided that $p_0(\tau)$ is narrowband with respect to $\mathrm{exp}(-\alpha_0\omega^2z)$, this solution can be approximated by $p_1(z,\tau) = p_0(\tau)\mathrm{exp}(-\alpha_0\omega_0^2z)$. Now, to second order:
\begin{equation}\label{EqSecondHarmonic}
    \frac{\partial p_2}{\partial z} - \alpha_0 \frac{\partial^2 p_2}{\partial \tau^2}= \frac{\beta}{\rho_\mathrm{L}c^3}p_0\frac{\partial p_0}{\partial \tau}e^{-2\alpha_0\omega_0^2z}.
\end{equation}
Here, the right hand has a center frequency $2\omega_0$ and acts as a source term for second-harmonic generation. 
With the boundary condition $p_2(0,\tau) = 0$, the solution to this equation is:
\begin{equation}
    p_2(z,\tau) = \mathcal{F}^{-1}\left\{ B(\omega)\frac{e^{-2\alpha_0\omega_0^2z}  - e^{-\alpha_0\omega^2z}}{\alpha_0\omega^2 - 2\alpha_0\omega_0^2}\right\}.
\end{equation}
with 
\begin{equation}
B(\omega) = \frac{\beta}{\rho_\mathrm{L}c^3}\mathcal{F}\left\{p_0\frac{\partial p_0}{\partial\tau}\right\}.
\end{equation}
For $p_0(\tau) \sim \mathrm{exp}(i\omega \tau)$ ($i = \sqrt{-1}$), this solution is the same as in~\cite{Blackstock1985}.

\subsection{Attenuated spherical wave}

For a medium with attenuation $\alpha(\omega) = \alpha_0\omega^2$, the wave equation can be written as (Szabo, second edition, Eq. 4.44,~\cite{Szabo2004}):
\begin{equation}
    \nabla^2v - \frac{1}{c^2}\frac{\partial^2 v}{\partial t^2} + \frac{2\alpha_0}{c}\frac{\partial v^3}{\partial t^3} = 0,
\end{equation}
where $v$ is the particle velocity, and $t$ is time. In spherical coordinates with spherical symmetry:
\begin{equation}
    \frac{1}{r^2}\frac{\partial}{\partial r}\left(r^2\frac{\partial v}{\partial r}\right) - \frac{1}{c^2}\frac{\partial^2 v}{\partial t^2} + \frac{2\alpha_0}{c}\frac{\partial v^3}{\partial t^3} = 0,
\end{equation}
where $v$ is the particle velocity.
With $\alpha_0\omega c << 1$, the harmonic solutions are (considering outward propagation waves only): $v(r,\tau) = A\mathrm{exp}(i\omega\tau-\alpha_0\omega^2r)/r$, with $\tau = t - r/c$ and $A$ a constant. With a small-amplitude approximation, the continuity equation can be written in spherical coordinates as:
\begin{equation}
    \frac{\partial p}{\partial t} = -\frac{\rho_\mathrm{L}c^2}{r^2}\frac{\partial (r^2 v)}{\partial r}.
\end{equation}
With $r>>c/\omega$ and $\alpha_0\omega c << 1$, we find $p = \rho_\mathrm{L}cv$.
Thus, the general (outward propagating) pressure wave can be written:
\begin{multline}
    p(r,\tau) = \int\frac{\hat{p}(\omega)}{r}e^{i\omega\tau - \alpha_0\omega^2r}d\omega = \\ 2\pi\mathcal{F}^{-1}\left\{\frac{\hat{p}(\omega)}{r}e^{- \alpha_0\omega^2r}\right\}
\end{multline}
for some function $\hat{p}(\omega)$. Let $p_s(r,\tau)$ be the solution in the absence of attenuation. For $r\rightarrow 0$, $p\rightarrow p_s$, and also, $p\rightarrow 2\pi\mathcal{F}^{-1}\{\hat{p}(\omega)/r\}$. Therefore, $2\pi\hat{p}(\omega)/r = \mathcal{F}\{p_s\}$. Thus, the attenuated solution can be written:
\begin{equation}
p(r,\tau) = \mathcal{F}^{-1}\{\mathcal{F}\{p_\mathrm{s}(r,\tau)\}\mathrm{exp}(-\alpha_0\omega^2r)\}.
\end{equation}

\subsection{Effect of data set size}\label{ResultsHyperparameter}

\begin{figure*}[ht]
    \centering
    \includegraphics[width=0.95\textwidth]{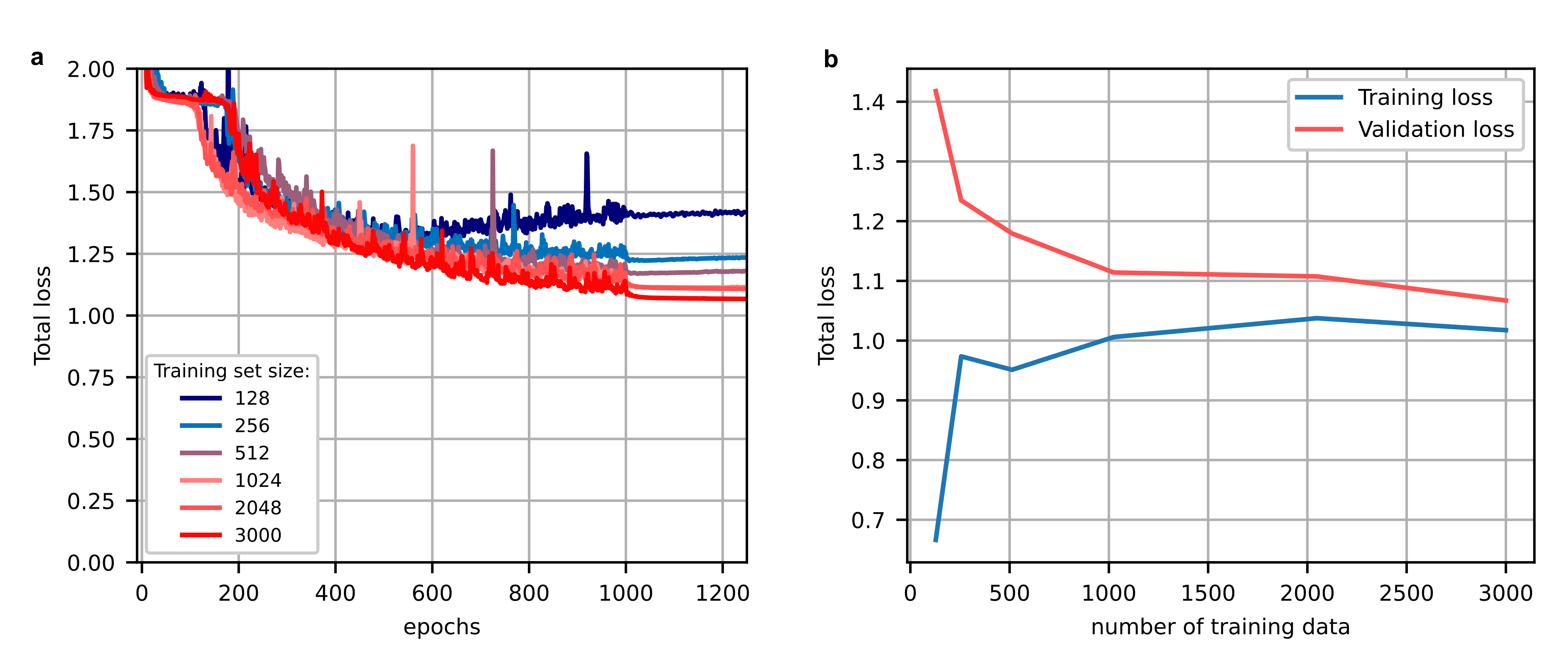}
    \caption{\textbf{Effect of data set size on model performance.} \textbf{a} Validation loss as a function of epochs for data sets of different sizes. \textbf{b} Final training loss and final validation loss as a function of data set size.}
    \label{blankeS2}
\end{figure*}

To determine what data set size is sufficient to train the network, we track the convergence of the loss for various training data set sizes (Fig.~\ref{blankeS2}a). The validation data set always contained 960 \textsc{rf} signals. The loss function settings are the same as for model HS$_{0.1}$. Fig.~\ref{blankeS2}b shows that the training loss and the validation loss converge to the same value after roughly 1000 epochs. Therefore, our models generalize well for unseen data within the domain of the simulation parameter space (acoustic pressure of 5-250 kPa, number of bubbles 10-1000).

\subsection{Simulation parameters}

Table~\ref{TableSimParam} gives the simulation parameters for the ultrasound simulator. The shell damping parameter $\kappa_s$ is computed with $\kappa_s=1.5\cdot10^{-9}\mathrm{exp}(8\cdot10^5R_0)$ (N$\cdot$s/m), where $R_0$ is the initial bubble radius in meters. This equation is a fit to the experimental data presented in Fig. 6B in~\cite{Segers2018}. The acoustic attenuation in dB/cm is $0.002f_\mathrm{MHz}^2$, where $f_\mathrm{MHz}$ is the frequency in MHz~\cite{Azhari2010}.

 \begin{table}[ht]
 \caption{Simulation parameters} 
\label{TableSimParam}
\centering
\begin{tabular}{@{\extracolsep{4pt}}llrl}
\toprule   
\textbf{Symbol}   & \textbf{Definition} & \textbf{Value} & \textbf{Unit}\\ 
\midrule
$c$ & Speed of sound in water at 20 $^{\circ}$C         & 1480 & m/s  \\ 
$\rho_\mathrm{L}$   & Density water at 20 $^{\circ}$C  & 998 & kg/m$^3$   \\ 
$\mu_\mathrm{vis}$ & Dynamic viscosity water at 20 $^{\circ}$C          & 0.0010 & Pa$\cdot$s \\ 
$P_0$ & Ambient pressure & 101.3 & kPa \\
$B/A$ & $B/A$ coefficient water & 5.2 & - \\
$\beta$ & Nonlinearity parameter & $1 + \frac{B}{2A}$ & -\\
\cmidrule{2-4}
 & \multicolumn{3}{c}{\textbf{Thermal model parameters}$^1$}\\
\cmidrule{2-4}
 & Equilibrium temperature & 293 & K \\
 & Molar mass C$_4$F$_{10}$  & 0.238 & kg/mol \\
 & Density C$_4$F$_{10}$  & 10.1 & kg/m$^3$ \\
 & Thermal conductivity water at 20 $^{\circ}$C & 0.6 & W/m/K\\ 
 & Thermal conductivity C$_4$F$_{10}$ & 0.0138 & W/m/K \\
 & Specific heat water at 20 $^{\circ}$C          & 4186 & in J/kg/K \\ 
 & Specific heat C$_4$F$_{10}$ & 809 & J/kg/K \\
 & Heat capacity ratio C$_4$F$_{10}$ & 1.07 & - \\
\bottomrule\\
\end{tabular}

\begin{tabular}{@{\extracolsep{4pt}}lcc}
\multicolumn{3}{p{251pt}}{$^1$These are not explicitely mentioned in the text. They are used in the model by Prosperetti~\cite{Prosperetti1977} to compute the effective polytropic exponent $\kappa$ and the effective thermal viscosity $\mu_\mathrm{th}$. }
\end{tabular}

\end{table}

\bibliographystyle{ieeetr}
\bibliography{References.bib}